\documentclass[fleqn,a4paper,times]{article}
\usepackage[latin9]{inputenc}
\usepackage{geometry}
\usepackage{array}
\usepackage{pifont}
\usepackage{bbding}
\usepackage{url}
\usepackage{multirow}
\usepackage{amsmath}
\usepackage{amsthm}
\usepackage{amssymb}
\usepackage{graphicx}
\usepackage[authoryear,round]{natbib}

\makeatletter

\providecommand{\tabularnewline}{\\}

\theoremstyle{definition}
\newtheorem{defn}{\protect\definitionname}
\theoremstyle{definition}
\newtheorem{condition}{\protect\conditionname}
\theoremstyle{plain}
\newtheorem{lyxalgorithm}{\protect\algorithmname}
\theoremstyle{remark}
\newtheorem{rem}{\protect\remarkname}

\@ifundefined{date}{}{\date{}}
\usepackage{microtype}
\usepackage{graphicx}
\usepackage{subfigure}
\usepackage{booktabs}
\usepackage{listings}
\usepackage{hyperref}
\usepackage{adjustbox}
\usepackage{arydshln}

\newcommand\dom{\mathrm{dom\,}}
\newcommand\ran{\mathrm{ran\,}}

\newcommand\calls{\rotatebox[origin=c]{180}{$\Lsh$}\,}

\makeatother

\usepackage{listings}
\providecommand{\algorithmname}{Algorithm}
\providecommand{\conditionname}{Condition}
\providecommand{\definitionname}{Definition}
\providecommand{\remarkname}{Remark}

\begin{document}
\title{Lazy object copy as a platform \\
for population-based probabilistic programming}
\author{Lawrence M. Murray\\
Uber AI}
\maketitle
\begin{abstract}
This work considers dynamic memory management for population-based
probabilistic programs, such as those using particle methods for inference.
Such programs exhibit a pattern of allocating, copying, potentially
mutating, and deallocating collections of similar objects through
successive generations. These objects may assemble data structures
such as stacks, queues, lists, ragged arrays, and trees, which may
be of random, and possibly unbounded, size. For the simple case of
$N$ particles, $T$ generations, $D$ objects, and resampling at
each generation, dense representation requires $O(DNT)$ memory, while
sparse representation requires only $O(DT+DN\log DN)$ memory, based
on existing theoretical results. This work describes an object copy-on-write
platform to automate this saving for the programmer. The core idea
is formalized using labeled directed multigraphs, where vertices represent
objects, edges the pointers between them, and labels the necessary
bookkeeping. A specific labeling scheme is proposed for high performance
under the motivating pattern. The platform is implemented for the
Birch probabilistic programming language, using smart pointers, hash
tables, and reference-counting garbage collection. It is tested empirically
on a number of realistic probabilistic programs, and shown to significantly
reduce memory use and execution time in a manner consistent with theoretical
expectations. This enables copy-on-write for the imperative programmer,
lazy deep copies for the object-oriented programmer, and in-place
write optimizations for the functional programmer.
\end{abstract}

\section{Introduction\label{sec:introduction}}

Probabilistic programming aims at better accommodating the workflow
of probabilistic modeling and inference in general-purpose programming
languages. Probabilistic programming languages (PPLs) support stochastic
control flow, allow probability distributions to be expressed over
rich data structures and higher-order functions, and are able to marginalize
and condition random variables using various inference techniques.

Among various approaches to the design and implementation of PPLs,
this work is most concerned with a category that might be called \emph{population-based}.
In a population-based\emph{ }PPL, multiple executions of a model are
maintained at any one time, via an inference method such as the particle
filter~\citep[see e.g.][]{Doucet2011} or Sequential Monte Carlo
(SMC) sampler~\citep{DelMoral2006}. PPLs with support for such methods
include LibBi~\citep{Murray2015}, Biips~\citep{Todeschini2014},
Venture~\citep{Mansinghka2014}, Anglican~\citep{Tolpin2016}, WebPPL~\citep{Goodman2014},
Figaro~\citep{Pfeffer2016}, and Turing~\citep{Ge2018}. The population-based
approach contrasts with, for example, gradient-based approaches that
make use of automatic differentiation to compute gradients for an
inference method such as Hamiltonian Monte Carlo, maximum likelihood
or variational inference. Some of the above languages support this
approach also. Other popular languages with support for gradient-based
inference include Stan~\citep{STAN} and Pyro~\citep{Bingham2019}.

Recent efforts have targeted tensor-based systems such as PyTorch~\citep{PyTorch}
and TensorFlow~\citep{TensorFlow}. PPLs such as Pyro~\citep{Bingham2019}
and Probabilistic Torch~\citep{Siddharth2017} use the former, Edward~\citep{Tran2017}
and TensorFlow Probability~\citep{Dillon2017} the latter. These
platforms are well-established for deep learning, offer dense representations
and automatic differentiation, and are well-suited to gradient-based
methods on models of bounded size. They are less suited to population-based
methods on models of potentially unbounded size, which may use stochastic
control flow in a way that causes multiple executions to diverge,
and be difficult to vectorize. Furthermore, they may use data structures
such as stacks, lists, trees, and ragged arrays, the size of which
may be unbounded and even random; they may grow or shrink as the program
executes and require dynamic memory allocation and deallocation. Tensor-based
platforms are not ideal for these data structures.

\begin{figure}[tp]
\begin{centering}
\includegraphics[width=0.7\columnwidth]{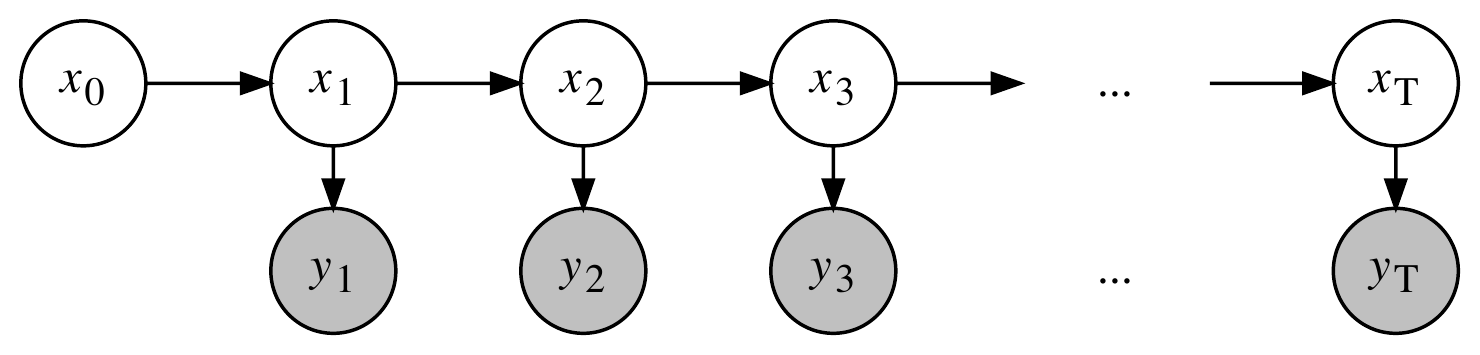}
\par\end{centering}
\caption{The state-space model as a directed graphical model, where arrows
represent conditional dependencies, unfilled nodes represent latent
variables, and filled nodes represent observed variables.\label{fig:ssm}}
\end{figure}

To better illustrate the memory usage pattern of a population-based
probabilistic program, consider a state-space model of $T$ time steps
with hidden states $x_{0:T}$ and observations $y_{1:T}$. The relationships
between these are represented mathematically via the joint distribution:
\[
p(x_{0:T},y_{1:T})=p(x_{0})\prod_{t=1}^{T}p(y_{t}\mid x_{t})p(x_{t}\mid x_{t-1}).
\]
The model is represented graphically in Figure \ref{fig:ssm}. We
will consider the different ways in which this model might be represented
\emph{programmatically}, i.e. as a computer program, below.

\begin{figure}[tp]
\begin{centering}
\includegraphics[width=0.7\columnwidth]{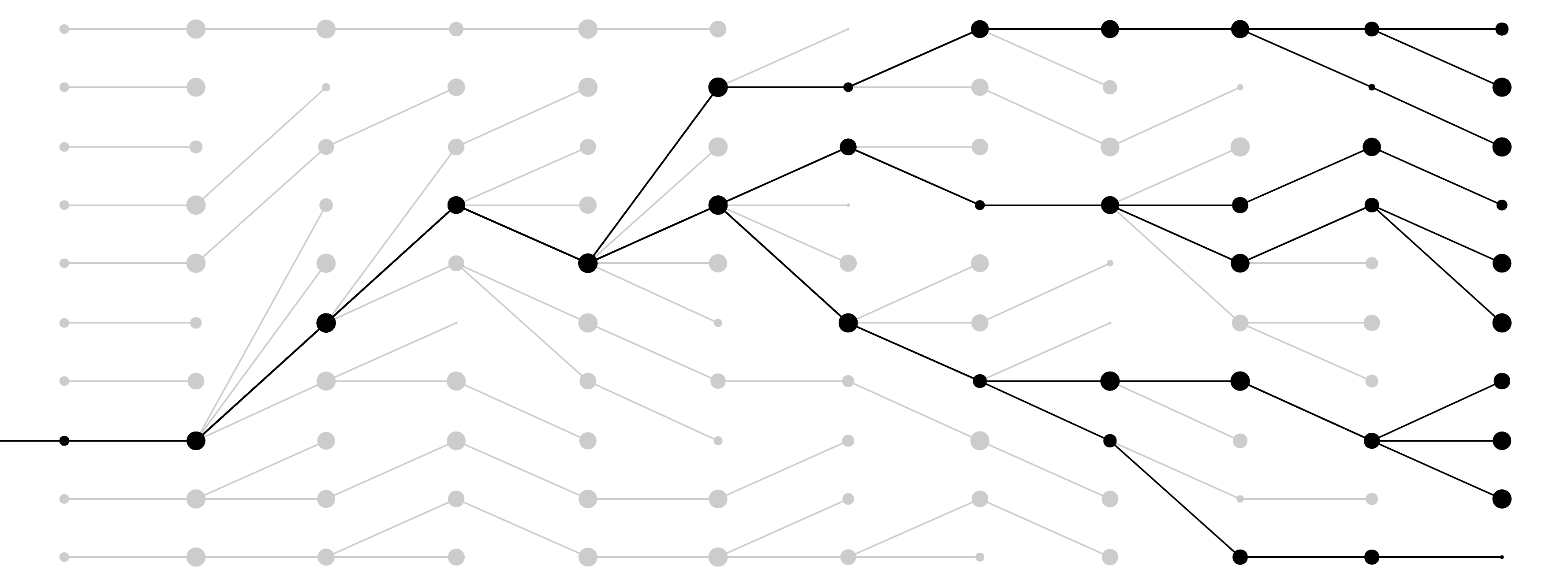}
\par\end{centering}
\caption{Example of a tree formed during execution of a particle filter. Time
$t=1,\ldots,T$ progresses from left to right. Each column represents
a population of particles $x_{t}^{1:N}$ as circles, their areas proportional
to $w_{t}^{1:N}$, each connected to its parent $x_{t-1}^{a_{t}^{n}}$
at the previous generation, chosen during resampling. The subtree
reachable from the final generation is shown in black, the remainder
in grey.\label{fig:tree}}
\end{figure}

A standard inference task for a PPL is, given observations $y_{1:T}$,
to compute the posterior distribution $p(x_{1:T}\mid y_{1:T})$. Various
approaches are available based on exact or approximate inference.
Consider the bootstrap particle filter~\citep{Gordon1993}, a building
block for more elaborate methods that include SMC as a generalization~\citep{DelMoral2006},
particle Markov chain Monte Carlo~\citep{Andrieu2010} methods, which
iterate a single filter, and SMC$^{2}$~\citep{Chopin2013} methods,
which run multiple filters simultaneously. Recent variational approaches
also use particle filters to estimate objectives~\citep{Maddison2017,Naesseth2018}.
The bootstrap particle filter maintains $N$ number of weighted \emph{particles},
each corresponding to an execution of the model:
\begin{enumerate}
\item For $n=1,\ldots,N$, initialize particle $x_{0}^{n}\sim p(x_{0})$
and weight $w_{0}^{n}=1$.
\item For $t=1,\ldots,T$ and $n=1,\ldots,N$:
\begin{enumerate}
\item Resample $a_{t}^{n}\sim\mathcal{C}(w_{t-1}^{1:N})$.
\item Propagate $x_{t}^{n}\sim p(x_{t}\mid x_{t-1}^{a_{t}^{n}})$.
\item Weight $w_{t}^{n}=p(y_{t}\mid x_{t}^{n})$.
\end{enumerate}
\end{enumerate}
Here, $\mathcal{C}(w_{t-1}^{1:N})$ represents the categorical distribution
on $\{1,\ldots,N\}$ with associated unnormalized probabilities $\{w_{t-1}^{1},\ldots,w_{t-1}^{N}\}$.
At each step $t$, the particles $x_{t}^{1:N}$ with weights $w_{t}^{1:N}$
approximate the distribution $p(x_{t}\mid y_{1:t})$. The particles
form a tree via the resampling step, as depicted in Figure \ref{fig:tree},
where for $t>0$ each particle $x_{t}^{n}$ has $x_{t-1}^{a_{t}^{n}}$
as its parent.

For many models, $x_{t}^{n}\in\mathbb{R}^{D}$, for $D$ number of
dimensions. A new generation is added at each time step, and previous
generations are immutable, such that $x_{1:T}^{1:N}$ may be stored
in a dense $D\times N\times T$ array (often called a tensor). To
reduce memory use it is preferable, however, to make use of the ancestor
indices $a_{1:T}^{1:N}$ to remove entries that are not reachable
from the latest generation. This is the case with LibBi~\citep{Murray2015},
for which the simple algorithm described and studied in \citet{Jacob2015}
was developed, which essentially uses a sparse representation for
the dimension of the array associated with $T$. \citet{Jacob2015}
establishes that the number of particles reachable from the $t$th
generation is bounded above by $t+cN\log N$, for some constant $c$.
This means that the dense approach requires $DNT$ storage, while
the sparse approach requires at most $DT+cDN\log DN$. These are often
interpreted for fixed $N$ and increasing $T$, where the memory saving
approaches the constant factor $N$. In practice, where $T$ is known
and an estimate of the likelihood $p(y_{1:T})$ is required, one typically
chooses $N=O(T)$ to control the variance of that estimate. One rule-of-thumb
to achieve this is to perform pilot runs with small $T$ and choose
an appropriate $N$ by trial and error; then, one gradually increases
$T$, and $N$ in proportion to it, until the full data set is accommodated
or computational limits reached. Taking this perspective, the memory
use for a task of size $T$ becomes more like $O(T^{2})$ for the
dense representation, reduced to $O(T\log T)$ for the sparse.

Models where $x_{t}^{n}$ is of some fixed type and size $D$ are
the simplest case. Amongst the models evaluated in Section~\ref{sec:evaluation},
$x_{t}^{n}$ includes stacks and ragged arrays of variable dimension,
binary trees, accumulators of sufficient statistics for variable elimination,
and mutation of previous states. All of these data structures are
better represented sparsely, or with multiple memory allocations that
can be separately deallocated, to enable the fastest algorithms for
adding, deleting and updating elements. The purpose of this work is
to easily enable such representations via a lazy object copy-on-write
platform. The platform must, of course, produce correct results, but
should also maximize object sharing, minimize object copying, and
minimize the bookkeeping required to achieve this. The motivating
usage pattern of tree-structured copies is used to direct design choices.

Immutable objects offer another paradigm. These are common in functional
programming languages, but also occur in imperative languages such
as Swift. Under this paradigm, in place of modifying an object, a
program creates a new object with the desired alteration. Because
this can greatly increase the number of object copies performed by
a program, good performance relies on either the use of persistent
data structures~\citep[see e.g.][]{Okasaki1999} that create the
appearance of incremental mutation without the computational expense,
or copy elimination and in-place mutation as a compiler optimization~\citep[see e.g.][]{Gopinath1989}.
In a sense, immutable objects pose the converse problem to that of
mutable objects: to determine when a copy is unnecessary, rather than
when a copy is necessary. While not the focus of the present work,
the proposed platform may be useful for this converse problem, allowing,
for example, a compiler to aggressively apply copy elimination and
in-place mutation, while deferring to runtime copy-on-write for correctness.

The operating system offers another alternative, as virtual memory
pages are typically copy-on-write in modern operating systems. \citet{Paige2014a}
make use of this facility, running particles as separate processes
that fork and terminate during resampling. The approach proposed in
the present work is finer-grained, working with individual objects
rather than pages, and a single process (with multiple threads) rather
than many. This increases the opportunity for sharing.

Section~\ref{sec:methods} provides a formal treatment of the ideas
using directed labeled multigraphs. This is intended to make the motivation
and correctness clear. In Section~\ref{sec:implementation} a concrete
implementation for the Birch PPL is sketched using smart pointers,
hash tables, and reference-counting garbage collection. In Section~\ref{sec:evaluation}
the approach is evaluated on a number of realistic probabilistic programs.
Section~\ref{sec:discussion} concludes.

\section{Methods\label{sec:methods}}

The proposed approach develops in three steps, defining:
\begin{enumerate}
\item $F$ as a directed multigraph, where vertices represent objects, and
edges the pointers between them, to demonstrate shallow and deep copies.
\item $G$ as a labeled directed multigraph that encodes one or more ongoing
\emph{lazy copies}, edges labeled with lists to identify them\emph{.}
\item $H$ as another labeled directed multigraph, edges now labeled with
single items to identify lazy copies within tree-structured patterns,
with eager copies outside of these patterns.
\end{enumerate}
Algorithms are provided to map $H$ to $G$ to $F$, making the equivalence
and correctness of the latter two representations clear.

\subsection{Shallow and deep copies}
\begin{defn}
Let $F=(V,E,s,t,b)$ be a directed multigraph, where:

\vspace{-3mm}
\begin{itemize}
\item $V$ is a set of vertices,
\item $E$ is a set of edges,
\item $s:E\rightarrow V$ maps each edge to its source vertex,
\item $t:E\rightarrow V$ maps each edge to its target vertex,
\item $b:V\rightarrow\cdot$ maps each vertex to its payload data, which
will be of some composite type as defined by a class; these types
need no be defined.
\end{itemize}
\end{defn}
Vertices represent objects and edges the pointers between them. One
particular vertex is the \emph{root} vertex, which, rather than representing
an object, encapsulates variables in global and local scopes. All
vertices are reachable from the root vertex by a directed path. If
a vertex becomes unreachable from the root vertex, it is assumed to
be deleted by garbage collection. Each vertex has some payload data
that does not require a formal specification, except to say that it
contains pointers, which establish the out-edges. Changes to these
pointers will change the out-edges accordingly, and so determine $s$
and $t$.

A \emph{copy} (qualified, if necessary, to \emph{shallow copy) }is
applied to an edge $e\in E$. The effect is to create a new vertex
$v$ with a copy of the payload data $b(t(e))$; this also copies
the out-edges of $t(e)$. A pointer is then created in some other
vertex $u$, creating a new edge $d$ with $s(d)=u$ and $t(d)=v$.

A \emph{deep copy} is likewise applied to an edge $e\in E$. The effect
is to copy the entire subgraph reachable from that edge, not just
the immediate target vertex of the edge. This proceeds recursively.
The first step is a shallow copy of $e$ to create a new vertex $v$,
the next step is a shallow copy applied to each out-edge of $v$,
and so on. The caveat is that each reachable vertex should be copied
only once. This requires keeping a record of objects that have already
been copied.

\begin{figure}[tp]
\begin{centering}
\hfill{}%
\begin{minipage}[t][1\totalheight][c]{0.15\columnwidth}%
\includegraphics[height=3cm]{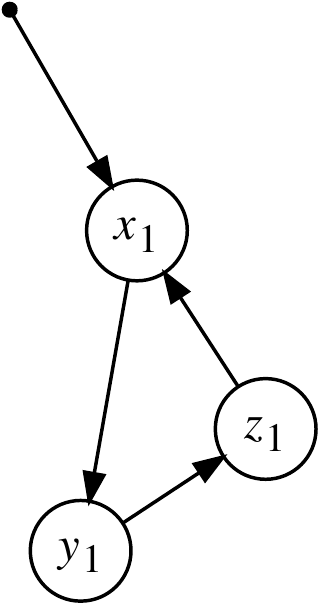}%
\end{minipage}\hfill{}%
\begin{minipage}[t][1\totalheight][c]{0.2\columnwidth}%
\includegraphics[height=3cm]{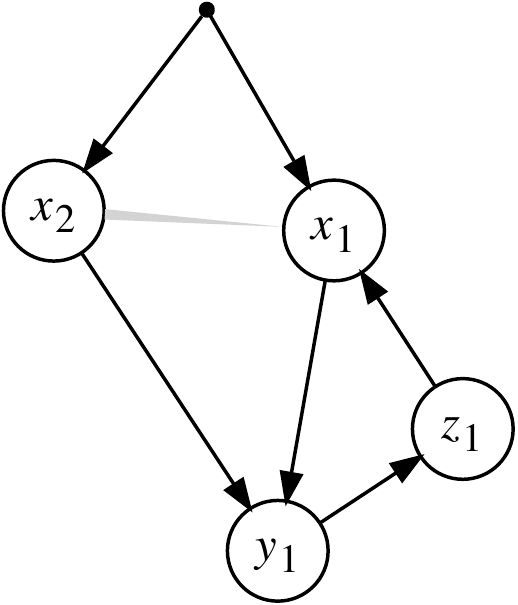}%
\end{minipage}\hfill{}%
\begin{minipage}[t][1\totalheight][c]{0.2\columnwidth}%
\includegraphics[height=3cm]{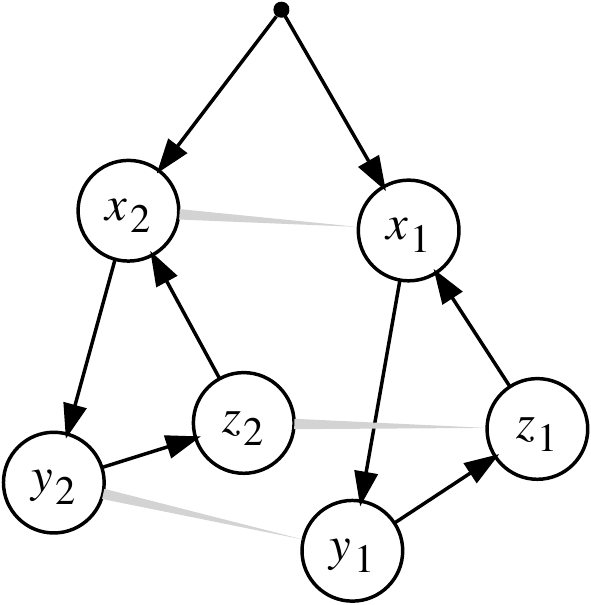}%
\end{minipage}\hfill{}%
\begin{minipage}[t][1\totalheight][c]{0.3\columnwidth}%
\begin{center}
\includegraphics[width=4.5cm]{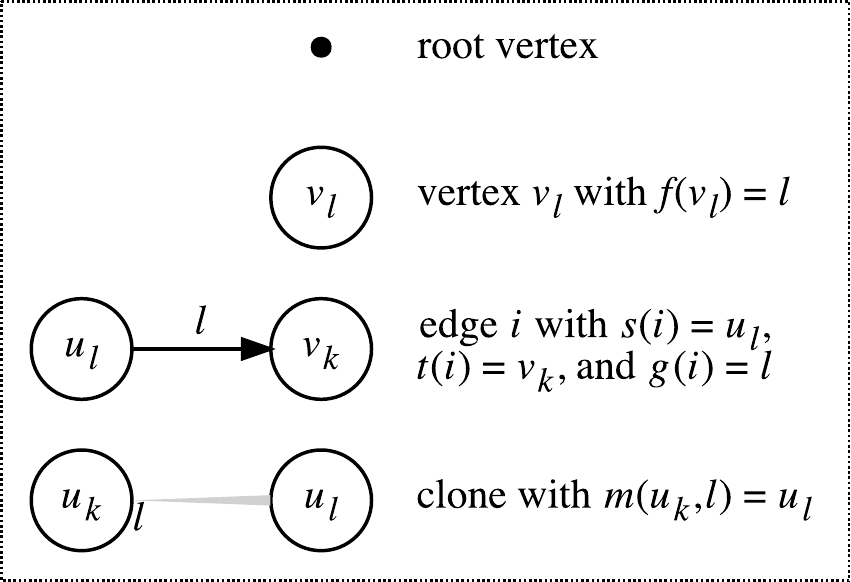}
\par\end{center}%
\end{minipage}
\par\end{centering}
\caption{Illustration of shallow and deep copy operations. On the left is the
original graph. In the middle is the result of applying a shallow
copy to the single out-edge of the root. On the right is the result
of applying a deep copy to the same edge.\label{fig:eager-deep-clone}}
\end{figure}

Figure \ref{fig:eager-deep-clone} illustrates the effect of shallow
and deep copy operations.

\subsection{Lazy copies}

\begin{figure}[tp]
\hfill{}\includegraphics[height=3cm]{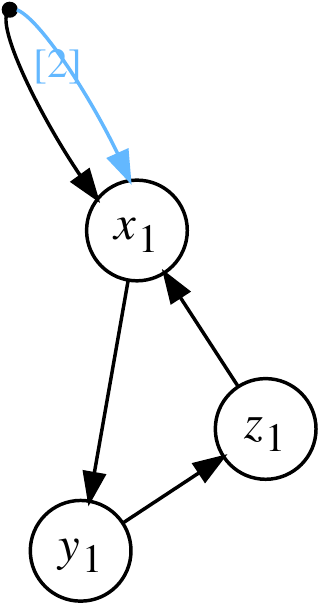}\hfill{}\includegraphics[height=3cm]{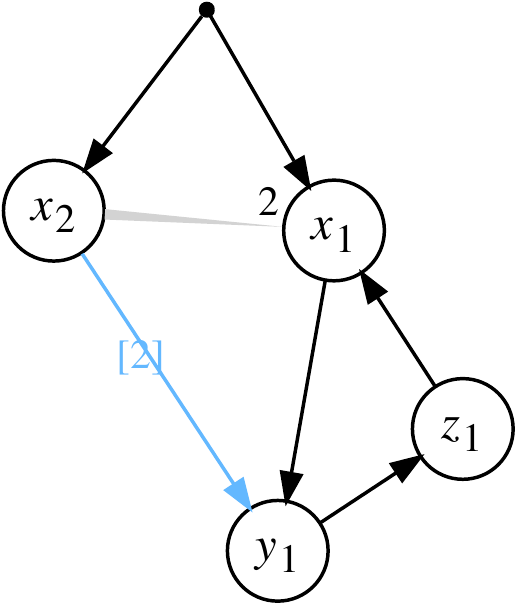}\hfill{}\includegraphics[height=3cm]{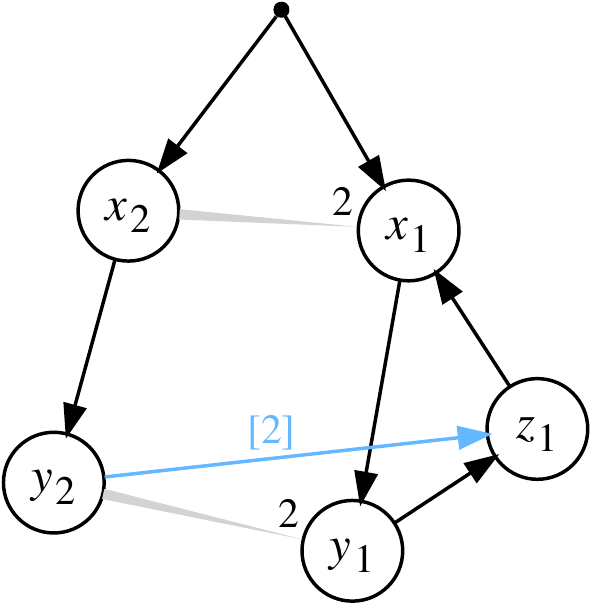}\hfill{}\includegraphics[height=3cm]{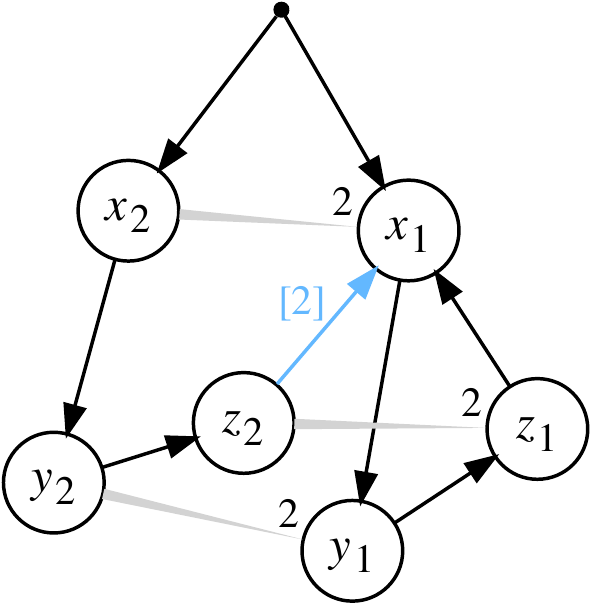}\hfill{}\includegraphics[height=3cm]{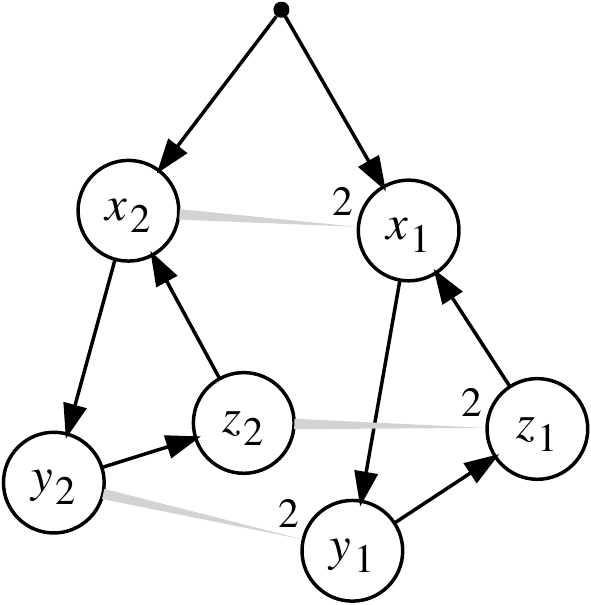}\hfill{}

\vspace{5mm}

\hfill{}\raisebox{1.4cm}{\includegraphics[height=1.6cm]{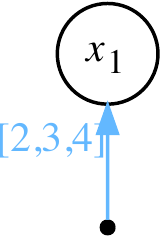}}\hfill{}\raisebox{1cm}{\includegraphics[height=2cm]{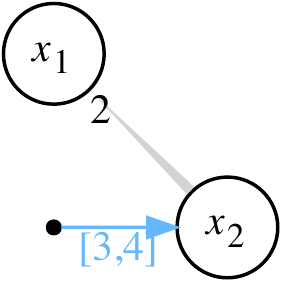}}\hfill{}\includegraphics[height=3cm]{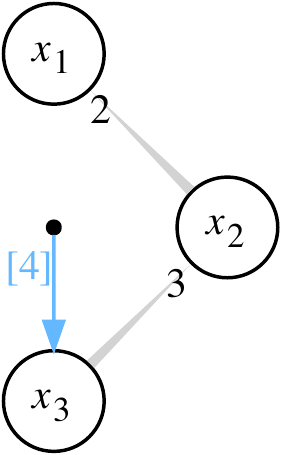}\hfill{}\includegraphics[height=3cm]{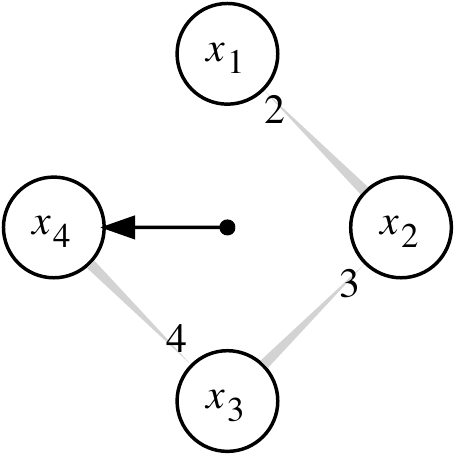}\hfill{}

\caption{Two illustrations of restoring $F$ from $G$ using Algorithm~\ref{alg:restore}.
From left to right, an edge with a non-empty label is chosen (shown
in color) and expanded to form the next state. \label{fig:lazy-deep-clone}}
\end{figure}

The aim is to preserve the appearance of a deep copy, while eliminating
unnecessary operations. Unnecessary copies occur when a deep copy
is performed, but not all vertices in the reachable subgraph will
need to be written, or even read, before they become unreachable.\emph{
Lazy copies} are proposed to achieve this, where multiple deep copies
can be underway concurrently, and vertices are only copied when a
write is necessary. This requires some bookkeeping, represented by
a labeling scheme on the graph.
\begin{defn}
Let $G=(V,E,s,t,b,R,L,m,f,g)$ be a labeled directed multigraph, where
$V$, $E$, $s$, $t$ and $b$ are as before, with new components:

\vspace{-3mm}
\begin{itemize}
\item $R\subseteq V$ is a set indicating read-only vertices,
\item $L$ is a set of labels, each of which identifies a distinct deep
copy operation,
\item $m:V\times L\nrightarrow V$ is a partial function that maps vertex
and label pairs to another vertex, providing a \emph{memo} of object
copies,
\item $f:V\rightarrow L$ assigns, to each vertex, a label identifying the
deep copy operation that created it,
\item $g:V\rightarrow[L]$ assigns, to each edge, a sequence of labels identifying
the deep copy operations through which the target vertex is yet to
be propagated.
\end{itemize}
\end{defn}
\begin{condition}
Any vertex $v\in V$ that has been copied has also been marked read-only,
i.e. for all $(v,l)\in\dom m$, $v\in R$.
\end{condition}
The graph $G$ is a compressed representation of $F$, combining identical
vertices. The additional components can be seen as the bookkeeping
necessary to restore $F$ from $G$. Algorithm~\ref{alg:restore}
provides the means to restore $F$ from $G$, Figure \ref{fig:lazy-deep-clone}
illustrates its application with examples.

\pagebreak{}
\begin{lyxalgorithm}
\label{alg:restore}Repeating the following procedure restores $F$
from $G$:

\vspace{-3mm}
\begin{enumerate}
\item Choose an an edge $e\in E$ such that $\left|g(e)\right|>0$ and $e$
is reachable from the root through a path in which each edge, $d$,
along that path has $\left|g(d)\right|=0$.
\item Let $v=t(e)$ and $l=\mathrm{head}\,g(e)$. If $(v,l)\in\dom m$ then
let $u=m(v,l)$, otherwise let $u$ be a copy of $v$ (i.e. $b(u)=b(v)$,
implying updates to $s$ and $t$ also) with $f(u)=l$. Update $t(e)\leftarrow u$
and $g(e)\leftarrow\mathrm{tail}\,g(e)$.
\end{enumerate}
\end{lyxalgorithm}
Algorithm~\ref{alg:restore} can be applied on demand to read or
write any desired vertex. Condition ~\ref{cond:read-or-write-vertex}
establishes the interface to make the graph $G$ look like $F$ from
the outside.
\begin{condition}
\label{cond:read-or-write-vertex}To read or write the data $b(v)$
of some vertex $v\in V$, that $v$ must be reachable from the root
through a path in which each edge $e$ has $\left|g(e)\right|=0$.
Algorithm~\ref{alg:restore} can be applied to achieve this.
\end{condition}

\subsection{Lazy copies on a tree}

The graph $G$ provides a straightforward way to reason about lazy
copies, but for implementation, labeling edges with lists adds significant
overhead. The motivating usage pattern suggests that copies tend to
be tree-structured, and that edge labels in $G$ will be paths on
this tree. Paths on trees can be represented more efficiently.
\begin{defn}
Let $H=(V,E,s,t,b,R,L,m,f,h,a)$ be a labeled directed multigraph
where all but the last two components are as for $G$. Function $g$
is replaced with the function $h$ and a new partial function $a$:

\vspace{-3mm}
\begin{itemize}
\item $h:V\rightarrow L$ assigns, to each edge, a single label,
\item $a:L\nrightarrow L$ is a partial function that assigns a parent to
each label other than a particular label designated the root, and
so arranging them in a tree.
\end{itemize}
\end{defn}
\begin{condition}
For all $e\in E$, there exists an $n\geq0$ such that $a^{n}(h(e))=f(t(e))$,
with the interpretation $a^{0}(l)=l$.

That is, the single label on each edge $e\in E$ is traced back through
$a$ until arriving at the label of its target vertex $f(t(e))$.
This restores a list that would be the equivalent label in $G$.
\end{condition}
\begin{lyxalgorithm}
To restore $G$ from $H$, for each $e\in E$:\vspace{-3mm}
\begin{itemize}
\item if $h(e)=f(t(e))$, $g(e)=[]$,
\item otherwise, set $g(e)=[a^{n-1}(h(e)),\ldots,a(h(e)),h(e)]$ where $a^{n}(h(e))=f(t(e))$.
\end{itemize}
\end{lyxalgorithm}
The labeling scheme of $H$ requires only a single label per edge,
which does not add much overhead, certainly compared to a list. It
is not as expressive as the labeling scheme of $G$, which can accommodate
arbitrary relationships between copy operations. The situations where
this generality is required are uncommon in the motivating usage pattern,
however, and when they occur, the proposed approach is to simply forego
the lazy copy and trigger an eager deep copy.

\subsection{Dynamics}

The graph $H$ represents only a snapshot in time of the memory state.
Of course, a program is dynamic, creating new vertices, writing data
to create and delete edges between them, and garbage collecting vertices
and edges that become unreachable. It is necessary to understand how
it evolves dynamically as a program executes. A number of new concepts
are necessary.
\begin{defn}
A thread of execution is considered to have a \emph{current context.
}This is a particular label $c\in L$, set as follows:

\vspace{-3mm}
\begin{enumerate}
\item At the start of the program, it is set to some arbitrary label, considered
the \emph{root context}. This is also the root of the tree defined
by $a$.
\item For some vertex $v\in V$, whenever its data $b(v)$ is being modified
or a member function (sometimes called a method) of $v$ is executing,
the current context is set to the label of that vertex, i.e. $c=f(v)$.
\end{enumerate}
\end{defn}
\begin{condition}
When a new vertex $v$ or new edge $e$ is created, its label is set
to the current context, i.e. $f(v)=c$ or $g(e)=c$.
\end{condition}
This means, for example, that if a member function of some vertex
$v\in V$ is executing, any new vertices that it creates along the
way are assigned the same label, $f(v)$. The current context can
be maintained using a stack, initialized with the root label, pushing
and popping labels as required by the above circumstances, and taking
the top of the stack as the current context at any time. In a multithreaded
environment, each thread requires its own context, and so maintains
its own stack.

For convenience, the memo $m$ is partitioned by labels $l\in L$,
where each $m_{l}$ contains all the entries of $m$ pertaining to
$l$ and all of its ancestors. Flattening in this way makes it unnecessary
to maintain the function $a$. We return to the motivation behind
this in Section~\ref{sec:implementation}.
\begin{defn}
\label{def:ml}For $l\in L$, define $m_{l}:V\rightarrow V:=\{u\mapsto v\in V\rightarrow V:\exists k\in a^{+}(l)((u,k)\mapsto v\in m)\}$,
where $a^{+}$ is the transitive closure of $a$.
\end{defn}
Below, pseudocode for operations on the graph $H$ is provided, with
worked examples of how real code triggers these operations as it executes.
The basic links between the two are that a deep copy operation in
real code, e.g.

\begin{lstlisting}
y <- deep_copy(x);
\end{lstlisting}

\vspace{-2mm}
will trigger a $\textsc{Deep-Copy}(x)$ operation, as defined in pseudocode.
Dereferencing an object to obtain the value of a member variable for
reading, e.g.

\begin{lstlisting}
value <- x.value;
\end{lstlisting}

\vspace{-2mm}
will trigger a $\textsc{Pull}(x)$ operation, while for writing, e.g.

\begin{lstlisting}
x.value <- value;
\end{lstlisting}

\vspace{-2mm}
will trigger a $\textsc{Get}(x)$ operation.

\pagebreak{}

As a simplification for pseudocode, assume that the sets $V$, $E$
and $L$ are automatically updated as objects are created, copied,
modified, and garbage collected. For example, when a new object $v$
is created, the update $V\leftarrow V\cup\{v\}$ is implied, and likewise,
if its payload data $b(v)$ is modified, $E$, $s$ and $t$ are updated
accordingly. For some function or partial function $\varphi$, the
notation $\varphi(x)\leftarrow y$ is used to insert a new entry $x\mapsto y$,
possibly replacing existing entries, i.e. $\varphi(x)\leftarrow y$
means exactly $\varphi\leftarrow\varphi\setminus\{x\mapsto\beta:\beta\in\ran\varphi\}\cup\{x\mapsto y\}$.

A deep copy is initialized on an edge:
\begin{lyxalgorithm}
~\vspace{-2mm}
\begin{align*}
 & \textsc{Deep-Copy}(e\in E)\\
 & \quad\textsc{Freeze}(e)\\
 & \quad\text{let }l\text{ be a new label}\\
 & \quad\text{set }m_{l}\leftarrow m_{h(e)}
\end{align*}
\end{lyxalgorithm}
The $\textsc{Freeze}$ is a recursive operation that marks reachable
vertices as read-only; it is defined later.

Read access to a vertex requires repeated application of Algorithm
\ref{alg:restore} to a single edge until it cannot be applied further
without creating a new vertex. This is encoded by the $\textsc{Pull}$
operation:
\begin{lyxalgorithm}
~\vspace{-2mm}
\begin{flalign*}
 & \textsc{Pull}(e\in E)\\
 & \quad\text{let }v=t(e)\\
 & \quad\text{let }l=h(e)\\
 & \quad\text{while }(v,l)\in\dom m\\
 & \quad\quad v\leftarrow m_{l}(v)\\
 & \quad\quad\text{update }t(e)\leftarrow v
\end{flalign*}
\end{lyxalgorithm}
After application, $t(e)$ targets the correct vertex for reading.
The pre-condition of Algorithm~\ref{alg:restore} is implicitly satisfied
by the fact that the program could not have discovered the edge $e$
without having first acquired its source vertex $s(e)$ via a $\textsc{Pull}$
operation, and so on recursively, back to the root vertex.

Write access to a vertex is encoded by the $\textsc{Get}$ operation,
which begins with $\textsc{Pull}$, but if the resulting vertex is
frozen, it must be copied before writing:
\begin{lyxalgorithm}
~\vspace{-2mm}
\begin{flalign*}
 & \textsc{Get}(e\in E)\\
 & \quad\textsc{Pull}(e)\\
 & \quad\text{let }v=t(e)\\
 & \quad\text{let }l=h(e)\\
 & \quad\text{if }v\in R\\
 & \quad\quad\text{let }u=\textsc{Copy}(e)\\
 & \quad\quad\text{update }t(e)\leftarrow u,m_{l}(v)\leftarrow u
\end{flalign*}
\end{lyxalgorithm}
After application, $t(e)$ targets the correct vertex for writing.
The operation $\textsc{Get}$ also has an implicit forwarding effect:
write attempts on a read-only vertex are forwarded to a copy.

An optimization is possible here. In many cases, there is only a single
in-edge to a vertex. There is no need to enter such vertices in $m_{l}$
when copied, as $m_{l}$ will never need to be queried for them.
\begin{rem}[Single-reference optimization]
\label{prop:single-reference-optimization}Let $v\in R$ be such
that:

\vspace{-3mm}
\begin{enumerate}
\item at the time of being frozen, the in-degree of $v$ is 1 and $v\notin\ran m$,
and
\item at the time of being copied, all the in-edges of $v$ have distinct
labels.
\end{enumerate}
When copying such a $v$, for any $l$, it is not necessary to update
$m_{l}$.
\end{rem}
The first condition requires not only that the in-degree of $v$ is
one, but also that it does not appear in the output of a memo. This
is because its appearance in a memo may compactly represent more than
one in-edge in the expanded graph. The condition is sufficient, but
not necessary, for the optimization to be valid. A sufficient and
necessary condition would require an expensive graph search. This
optimization is particularly important, and its effect evaluated separately
in Section~\ref{sec:evaluation}.

The $\textsc{Get}$ operation requires a shallow copy operation, which
proceeds as follows:
\begin{lyxalgorithm}
\label{alg:copy}~\vspace{-2mm}
\begin{flalign*}
 & \textsc{Copy}(e\in E)\rightarrow V\\
 & \quad\text{let }v=t(e)\\
 & \quad\text{for }d\in\{i\in E\mid s(i)=v\}\\
 & \quad\quad\text{if }h(d)\neq f(v)\\
 & \quad\quad\quad\textsc{Finish}(d)\text{}\\
 & \quad\quad\quad\textsc{Freeze}(d)\\
 & \quad\text{let }u\text{ be a new vertex with }b(u)=b(v)\\
 & \quad\text{return }u
\end{flalign*}
\end{lyxalgorithm}
The conditional triggers an eager deep copy for usage patterns outside
that of a tree. This occurs when copying some edge $d\in E$ where
$h(d)\neq f(v)$. We call such edges \emph{cross references}. Cross
references are precisely those that are outside the tree-structured
usage pattern, and can be represented in the labeling scheme of $G$,
but not $H$.

The two operations $\textsc{Finish}$ and $\textsc{Freeze}$ have
similar recursive structures. The first ensures that all vertices
in the subgraph reachable from an edge are read-only, i.e. put in
the set $R$:
\begin{lyxalgorithm}
~\vspace{-2mm}
\begin{flalign*}
 & \textsc{Freeze}(e\in E)\\
 & \quad\text{let }v=t(e)\\
 & \quad\text{if }v\notin R\\
 & \quad\quad R\leftarrow R\cup\{v\}\\
 & \quad\quad\text{for }d\in\{i\in E\mid s(i)=v\}\\
 & \quad\quad\quad\textsc{Freeze}(d)
\end{flalign*}
\end{lyxalgorithm}
The second ensures that all lazy copies in the subgraph have been
completed:
\begin{lyxalgorithm}
~\vspace{-2mm}
\begin{flalign*}
 & \textsc{Finish}(e\in E)\\
 & \quad\text{let }v=t(e)\\
 & \quad\text{if }h(e)\neq f(v)\\
 & \quad\quad\textsc{Get}(e)\\
 & \quad\quad\text{for }d\in\{i\in E\mid s(i)=v\}\\
 & \quad\quad\quad\textsc{Finish}(d)
\end{flalign*}
\end{lyxalgorithm}
The intent is that real code will trigger the above operations, given
in pseudocode, as it executes. Consider, first, the following class,
intended to serve as the type of nodes in a singly-linked list of
integers:\texttt{\small{}}
\begin{lstlisting}
class Node {
  value:Integer;
  next:Node;
}
\end{lstlisting}
The member variables \texttt{value} and \texttt{next} hold the value
of the current node and a pointer to the next node, respectively.
Assume that variables of primitive types like \texttt{Integer} are
kept as values, while variables of class types like \texttt{Node}
hold pointers to those objects, dynamically allocated.

This basic class is sufficient to explain how real code triggers the
pseudocode operations. Table \ref{tab:standard-example} provides
an example of the standard usage pattern, where deep copy operations
relate as a tree. Table \ref{tab:nonstandard-example} gives an example
outside of the standard usage pattern, where there exists a cross
reference, and an incorrect result would be produced if not for the
special treatment of cross references in Algorithm~\ref{alg:copy}.

\begin{table}[tp]
\begin{tabular}[t]{>{\raggedright}p{0.28\textwidth}>{\raggedright}p{0.15\textwidth}>{\raggedright}p{0.17\textwidth}>{\raggedright}p{0.3\textwidth}}
\textbf{Code} & \textbf{Operations} & \textbf{State} & \textbf{Commentary}\tabularnewline
\hline 
\texttt{\small{}x1:Node;} &  & \multirow{5}{0.17\textwidth}{\includegraphics[scale=0.4]{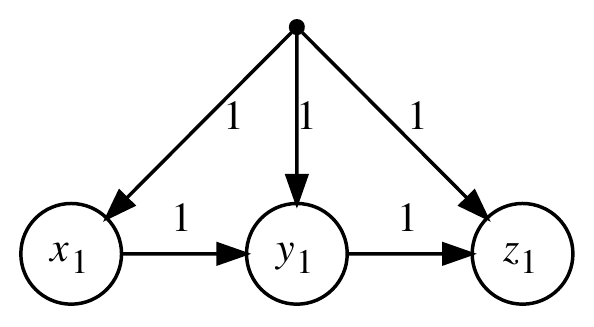}} & \tabularnewline
\texttt{\small{}y1:Node;} &  &  & \tabularnewline
\texttt{\small{}z1:Node;} &  &  & \tabularnewline
\texttt{\small{}x1.next \textless - y1;} &  &  & \tabularnewline
\texttt{\small{}y1.next \textless - z1;} &  &  & \tabularnewline
\hdashline[1pt/1pt]

\texttt{\small{}x2:Node \textless - deep\_copy(x1);} & $\textsc{Deep-Copy}(x_{1})$ & \multirow{1}{0.17\textwidth}{\includegraphics[scale=0.4]{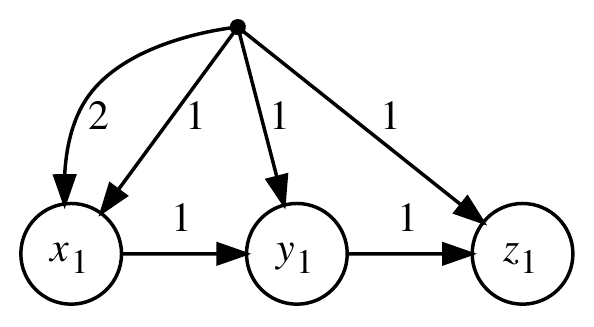}} & \multirow{1}{0.3\textwidth}{A new label 2 is created, and a new edge, but no new vertex.}\tabularnewline
 &  &  & \tabularnewline
 &  &  & \tabularnewline
 &  &  & \tabularnewline
\hdashline[1pt/1pt]

\texttt{\small{}value \textless - x2.value;} & $\textsc{Pull}(x_{2})$ &  & Read-only access, copy not required, state unchanged.\tabularnewline
\hdashline[1pt/1pt]

\texttt{\small{}x2.value \textless - value;} & $\textsc{Get}(x_{2})$ & \multirow{2}{0.17\textwidth}{\includegraphics[scale=0.4]{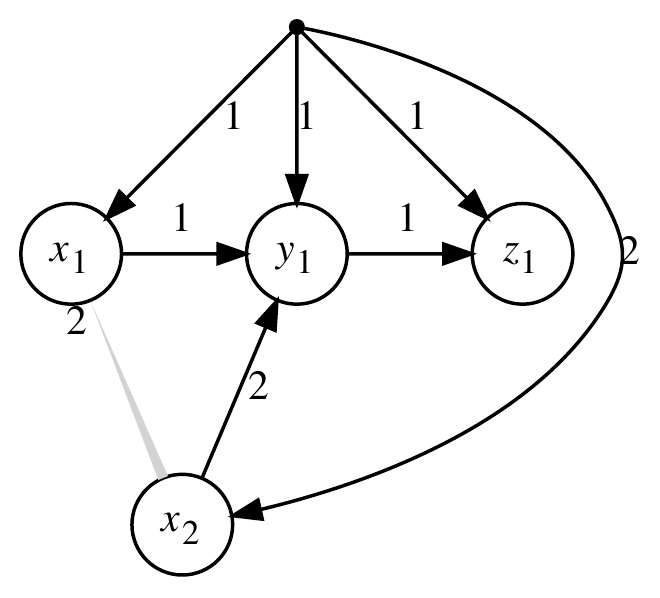}} & \multirow{2}{0.3\textwidth}{Write access, copy required.}\tabularnewline
 & $\quad\calls\textsc{Copy}(x_{1})$ &  & \tabularnewline
 &  &  & \tabularnewline
 &  &  & \tabularnewline
 &  &  & \tabularnewline
 &  &  & \tabularnewline
\hdashline[1pt/1pt]

\texttt{\small{}y2:Node \textless - x2.next;} & $\textsc{Get}(x_{2})$ & \multirow{4}{0.17\textwidth}{\includegraphics[scale=0.4]{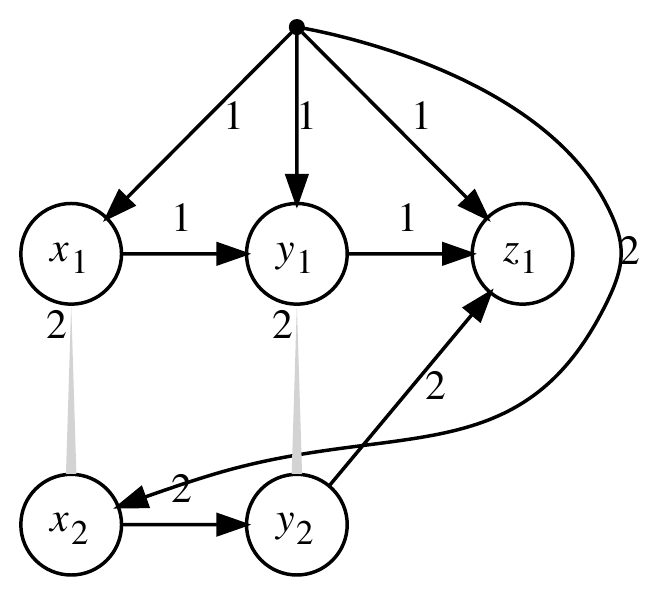}} & \multirow{4}{0.3\textwidth}{As each node in the list is accessed it must be copied, as write access
is potentially required to update its \texttt{\small{}next} pointer.}\tabularnewline
\texttt{\small{}z2:Node \textless - y2.next;} & $\textsc{Get}(y_{2})$ &  & \tabularnewline
 & $\quad\calls\textsc{Copy}(y_{1})$ &  & \tabularnewline
 &  &  & \tabularnewline
 &  &  & \tabularnewline
 &  &  & \tabularnewline
\hdashline[1pt/1pt]\texttt{\small{}value \textless - z2.value;} & $\textsc{Pull}(z_{2})$ &  & Read-only access, copy not required, state unchanged.\tabularnewline
\hdashline[1pt/1pt]

\texttt{\small{}z2.value \textless - value;} & $\textsc{Get}(z_{2})$ & \multirow{4}{0.17\textwidth}{\includegraphics[scale=0.4]{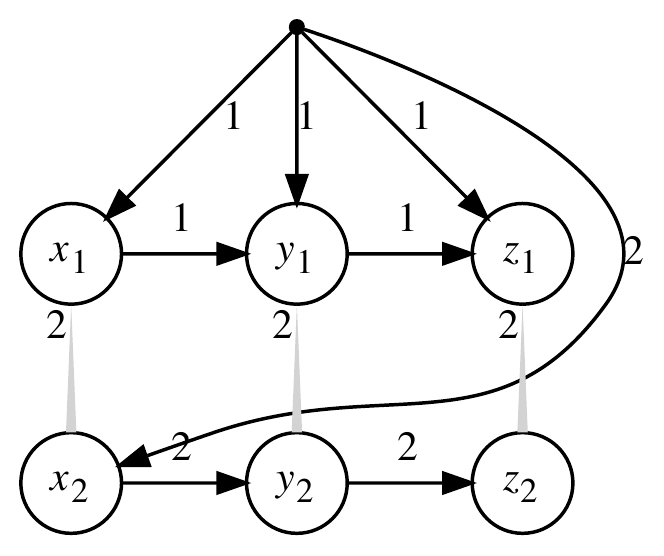}} & \multirow{4}{0.3\textwidth}{Write access, copy required.}\tabularnewline
 & $\quad\calls\textsc{Copy}(z_{1})$ &  & \tabularnewline
 &  &  & \tabularnewline
 &  &  & \tabularnewline
 &  &  & \tabularnewline
 &  &  & \tabularnewline
\hline 
\end{tabular}

\caption{Example of the standard use case of several lazy deep copy operations
related as a tree. As the program in the first column executes it
triggers the operations indicated in the second column, updating the
state of the multigraph depiction of the memory state in the third
column.\label{tab:standard-example}}
\end{table}

\begin{table}[tp]
\begin{tabular}{>{\raggedright}p{0.28\textwidth}>{\raggedright}p{0.2\textwidth}>{\raggedright}p{0.12\textwidth}>{\raggedright}p{0.3\textwidth}}
\textbf{Code} & \textbf{Operations} & \textbf{State} & \textbf{Commentary}\tabularnewline
\hline 
\texttt{\small{}x1:Node;}{\small\par}

\texttt{\small{}x1.value \textless - 1;} &  & \multirow{2}{0.12\textwidth}{\includegraphics[scale=0.5]{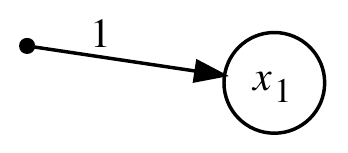}} & \tabularnewline
 &  &  & \tabularnewline
\hdashline[1pt/1pt]

\texttt{\small{}x2:Node \textless - deep\_copy(x1);} & $\textsc{Deep-Copy}(x_{1})$ & \multirow{1}{0.12\textwidth}{\includegraphics[scale=0.5]{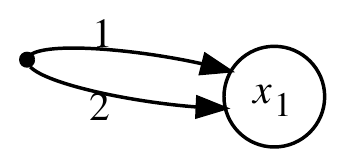}} & \tabularnewline
 &  &  & \tabularnewline
\hdashline[1pt/1pt]

\texttt{\small{}x2.value \textless - 2;} & $\textsc{Get}(x_{2})$ & \multirow{2}{0.12\textwidth}{\includegraphics[scale=0.5]{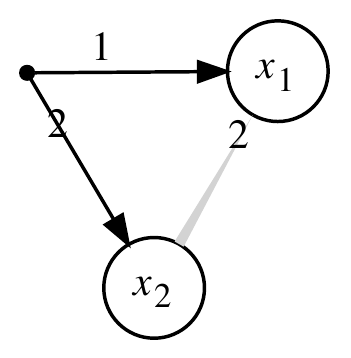}} & \tabularnewline
 & $\quad\calls\textsc{Copy}(x_{1})$ &  & \tabularnewline
 &  &  & \tabularnewline
 &  &  & \tabularnewline
 &  &  & \tabularnewline
\hdashline[1pt/1pt]

\texttt{\small{}x2.next \textless - x1;} &  & \multirow{2}{0.12\textwidth}{\includegraphics[scale=0.5]{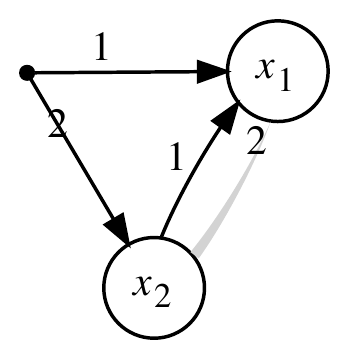}} & \multirow{2}{0.3\textwidth}{This establishes a cross reference: the vertex $x_{2}$ has an outgoing
edge with a different label ($=1$) to itself ($=2$).}\tabularnewline
 &  &  & \tabularnewline
 &  &  & \tabularnewline
 &  &  & \tabularnewline
 &  &  & \tabularnewline
\hdashline[1pt/1pt]\texttt{\small{}x3:Node \textless - deep\_copy(x2);} & $\textsc{Deep-Copy}(x_{2})$ & \multirow{2}{0.12\textwidth}{\includegraphics[scale=0.5]{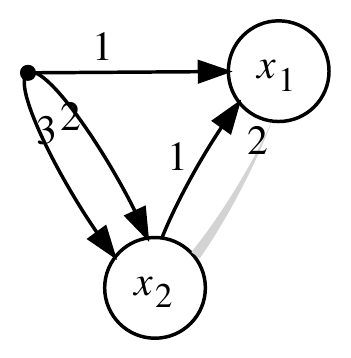}} & \tabularnewline
 &  &  & \tabularnewline
 &  &  & \tabularnewline
 &  &  & \tabularnewline
 &  &  & \tabularnewline
\hdashline[1pt/1pt]

\texttt{\small{}x3.value \textless - 3;} & $\textsc{Get}(x_{3})$ & \multirow{6}{0.12\textwidth}{\includegraphics[scale=0.5]{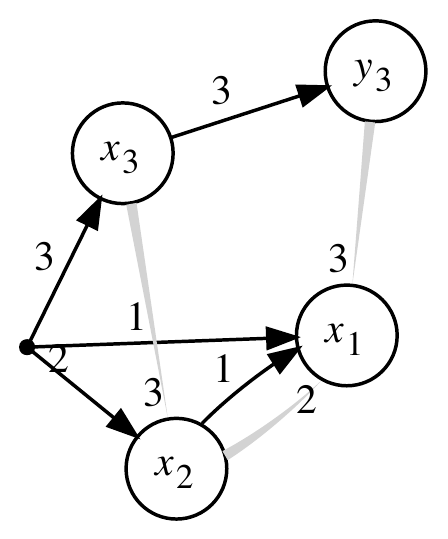}} & \textbf{\Checkmark{} Correct}\tabularnewline
 & $\quad\calls\textsc{Copy}(x_{1})$ to $y_{3}$ &  & \tabularnewline
 & $\quad\quad\calls\textsc{Finish}(y_{3})$ &  & \tabularnewline
 &  &  & \tabularnewline
\texttt{\small{}y3:Node \textless - x3.next;} &  &  & \tabularnewline
\texttt{\small{}print(y3.value);} & $\textsc{Pull}(y_{1})$ &  & Prints 1, which is correct.\tabularnewline
 &  &  & \tabularnewline
\hdashline[1pt/1pt]

\texttt{\small{}x3.value \textless - 3;} & $\textsc{Get}(x_{3})$ & \multirow{4}{0.12\textwidth}{\includegraphics[scale=0.5]{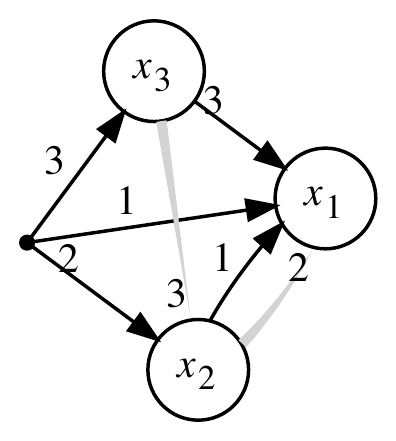}} & \textbf{\ding{56} Incorrect}\tabularnewline
 & $\quad\calls\textsc{Copy}(x_{2})$ &  & \tabularnewline
\texttt{\small{}y3:Node \textless - x3.next;} &  &  & \tabularnewline
\texttt{\small{}print(y3.value);} & $\textsc{Pull}(y_{1})$ &  & Prints 2, which is incorrect. This is because the edge $x_{3}\rightarrow x_{1}$
has label 3, implying a list label of $[2,3]$ under graph style $G$,
not $[3]$ as in the former (correct) case. This misdirects the pointer
to the wrong vertex when expanded.\tabularnewline
 &  &  & \tabularnewline
\hline 
\end{tabular}

\caption{Example of a use case outside the standard, where assignment creates
a cross reference, and a $\textsc{Finish}$ operation is triggered
for correctness. The second last row depicts the correct behaviour,
the last row depicts a counterfactual that would produce an incorrect
result.\label{tab:nonstandard-example}}
\end{table}

\section{Implementation\label{sec:implementation}}

The proposed approach has been implemented for Birch~\citep{Murray2018a},
an imperative and object-oriented language with copyable (multi-shot)
coroutines and specialized operators that provide ergonomic support
for population-based probabilistic programming. Objects are allocated
on the heap and pass by reference, with reference counting garbage
collection. Primitive values pass by value. Arrays are semantically
primitive values that pass-by-value, but are implemented on the heap
with a separate copy-on-write mechanism for their simpler requirements.
The copy-on-write mechanism described in this work is intended for
the more complex case of objects; this includes the copyable coroutines
in Birch, the closures of which are implemented as objects with mutable
state. Birch compiles to C++. The platform described in this work
has been implemented in the C++ library that supports this generated
code.

The graph formalism is translated as follows:
\begin{itemize}
\item Each label $l\in L$ becomes an object, which keeps its associated
memo $m_{l}$ as a hash table of pointers.
\item Each vertex $v\in V$ becomes an object, which keeps, among its data
$b(v)$, a pointer to the object representing its label, $f(v)$.
\item Each edge $e\in E$ becomes a pair of pointers among the data of its
source vertex $s(e)$. The first pointer is to the object representing
the vertex $t(e)$, the second to the object representing the label
$h(e)$. These pairs are referred to as \emph{lazy pointers}.
\end{itemize}
In the implementation suggested here, the pointers are smart pointers
that maintain shared and weak reference counts on objects to support
a reference-counting garbage collector. With some adaptations, a tracing
garbage collector could be used instead.

Because labels are implemented as separate objects containing memos,
and lazy pointers are pairs of smart pointers referencing an object
and a label, extra reference cycles are introduced that would not
otherwise exist. To eliminate these, a vertex $v$ does not increment
the reference count on its label $f(v)$, and any lazy pointers $e$
amongst its members only increment the reference count on their label
if they are a cross reference, i.e. if $h(e)\neq f(v)$. This breaks
the additional reference cycles.

Another issue is that memos may keep objects reachable longer than
necessary. This occurs when the only reference to an object is a key
in the hash table for a memo. To collect these, a new reference count
is introduced for each object: a \emph{memo }count alongside the regular
shared and weak counts. Keys in the hash tables increment the memo
count only. A sweep of a table can be performed at any point to remove
entries with zero shared and weak count, but nonzero memo count. These
sweeps occur when resizing and copying hash tables. The three reference
counts work together as follows:
\begin{enumerate}
\item A new object is initialized with shared, weak, and memo counts of
one.
\item When the shared count reaches zero the object is destroyed and the
weak count decremented by one.
\item When the weak count reaches zero the memo count is decremented by
one.
\item When the memo count reaches zero, memory is freed.
\end{enumerate}
Reference counting facilitates the single-reference optimization described
in Remark~\ref{prop:single-reference-optimization}. The chosen implementation
is to flag an object that meets the condition at time of freezing.
It is allowed that reference counts on the object subsequently change,
as long as each in-edge has a distinct label. This is violated only
if an assignment occurs that creates an identical in-edge. In this
situation $\textsc{Get}$ is triggered on the edge, maintaining distinct
labels.

The single-reference optimization avoids the update of a memo according
to an object having only a single reference \emph{at the time of being
frozen}. This is different to copy elimination for an object having
only a single reference \emph{at the time of being copied}. This latter
optimization is also used. It requires that a frozen object can be
\emph{thawed} for reuse.

Although not detailed here, atomic operations are used judiciously
to ensure thread safety.

Finally, the choice to use $m_{l}$, given in Definition \ref{def:ml},
requires some justification. As well as simplifying the implementation,
this is justified by cache efficiency. There are a number of choices:
a single memo $m:V\times L\rightarrow V$, partitioning by vertex
to keep a separate memo $m_{v}:L\rightarrow V$ for each vertex $v\in V$,
or partitioning by label to keep a separate memo $m_{l}:V\rightarrow V$
for each label $l\in L$. Access to the memo is expected to have temporal
locality with respect to label, i.e. with high probability a query
$m(u,l)$ will be followed by a query $m(v,l)$ for different vertices
$u$ and $v$ but identical label $l$. This $l$ will often be the
current context, $c$. If $m$ is implemented as a hash table, the
second query may benefit from some part of that hash table being drawn
into cache by the first query. Consequently, the partition by label,
$m_{l}$, is preferred. A further choice is whether or not to explicitly
maintain the tree, via the partial function $a$, or to flatten it.
Anecdotally, the flattening has produced a simpler and faster implementation,
and is recommended.

\section{Evaluation\label{sec:evaluation}}

The implementation is evaluated with regards to execution time and
memory use across a number of realistic probabilistic models and inference
methods. Each test case is run for two tasks:
\begin{enumerate}
\item performing inference with a given data set, and
\item performing simulation with no data set.
\end{enumerate}
No copies occur in the second task; its purpose is to isolate the
additional resource requirements of lazy pointers. For each task there
are three configurations:
\begin{enumerate}
\item eager copy,
\item lazy copy, and
\item lazy copy with the single-reference optimization.
\end{enumerate}
These are compile-time configurations that compile away unnecessary
bookkeeping. The problems are as follows:
\begin{enumerate}
\item \textbf{RBPF }A mixed linear-nonlinear state-space model as described
in \citet{Lindsten2010}, with a Rao--Blackwellized particle filter
via delayed sampling~\citep{Murray2018}, with $N=2048$ and $T=500$
for both inference and simulation.
\item \textbf{PCFG }A probabilistic context-free grammar model (unpublished)
with an auxiliary particle filter~\citep{Pitt1999} with custom proposal,
$N=16384$, $T=3262$ for inference, and $T=2000$ for simulation.
\item \textbf{VBD }A vector-borne-disease model as described in \citet{Murray2018},
with dengue data set from Micronesia as in~\citet{Funk2016a}, with
a marginalized particle Gibbs method as described in \citet{Wigren2019}
for 3 iterations, $N=4096$, $T=182$ for inference, and $T=400$
for simulation. With this method, there is a deep copy of a single
particle between iterations that must be completed eagerly, as it
is outside the tree pattern.
\item \textbf{MOT }A multi-object tracking model for an unknown number of
objects with linear-Gaussian dynamics, as described in \citet{Murray2018a},
with simulated data, $N=4096$, $T=100$ for inference, and $T=300$
for simulation.
\item \textbf{CRBD }A constant rate birth-death model over a tree with an
alive particle filter~\citep{DelMoral2015b} and delayed sampling
as described in \citet{Kudlicka2019}, with data set of cetacean phylogeny~\citep{Steeman2009},
$N=5000$ and $T=173$ for both inference and simulation.
\end{enumerate}
All combinations of task, configuration, and problem are compiled
and run for 20 repetitions, on a system with two Intel Xeon E5-2680
CPUs (each 12 cores, 30 MiB cache) and 64 GiB memory. The C++ compiler
used is GCC 9.1.0. The propagation and weighting of particles is parallelized
across 24 threads, one bound to each hardware core, using OpenMP with
static scheduling. Random number seeds are matched across configurations,
using a different seed for each repetition. For each task and problem,
the output is expected to match regardless of the configuration; a
comparison of output files confirms that this is the case.

\begin{figure*}[t]
\includegraphics[width=0.5\textwidth]{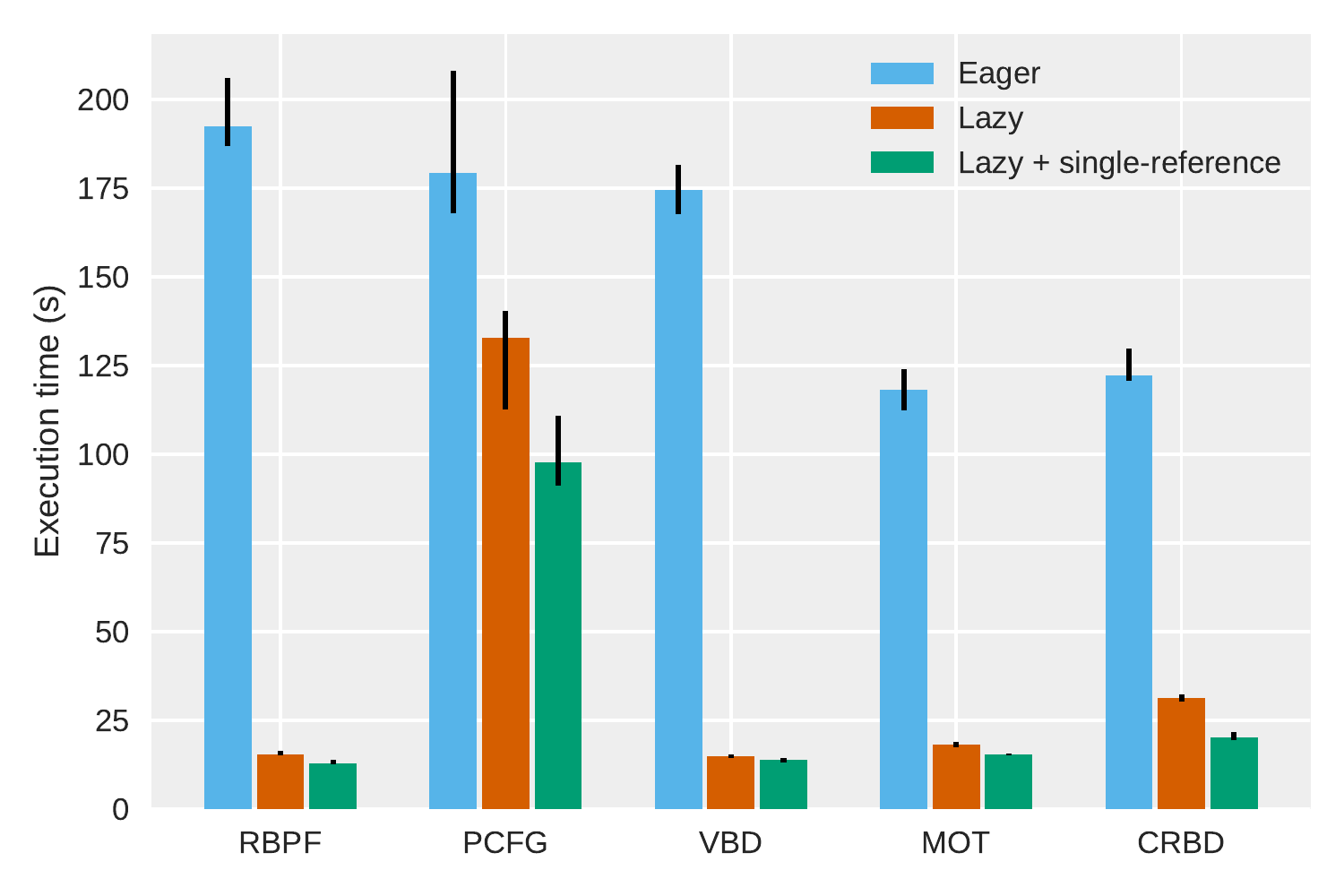}\includegraphics[width=0.5\textwidth]{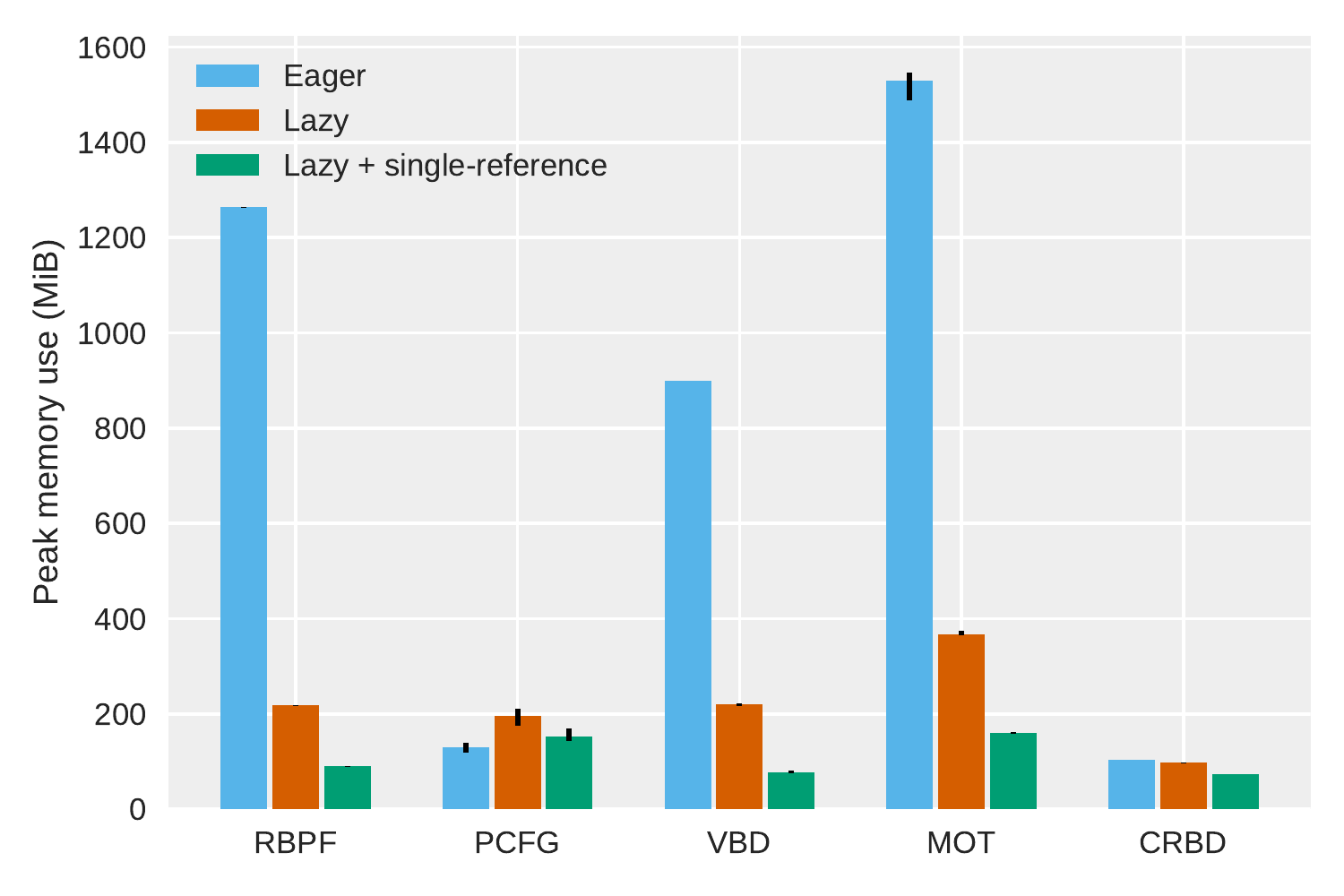}

\caption{Execution time (left) and peak memory use (right) for the inference
task. Heights indicate median, and error bars the interquartile range,
across 20 runs.\label{fig:inference}}
\end{figure*}

Figure~\ref{fig:inference} reports execution time and peak memory
use. On the inference task (with the exception of the PCFG problem,
discussed below), significant reductions are apparent when using lazy
copies compared to eager copies, with even greater reductions when
the single-reference optimization is applied. For the simulation task,
comparable execution time and increased memory use is apparent, as
expected. This additional memory use is attributed to an extra 8 bytes
per pointer and 12 bytes per object to support lazy copies, even if
unused.

\begin{figure*}[t]
\includegraphics[width=0.5\textwidth]{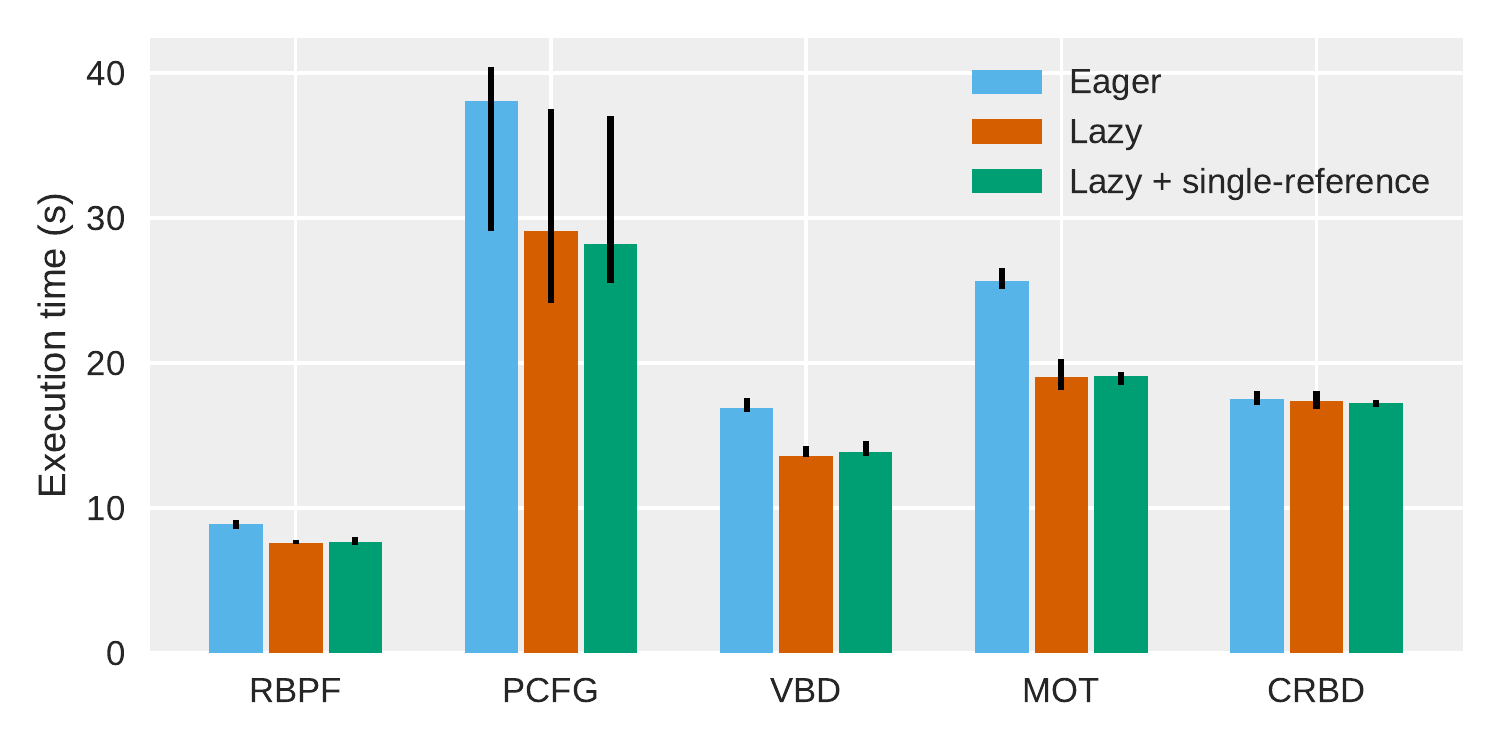}\includegraphics[width=0.5\textwidth]{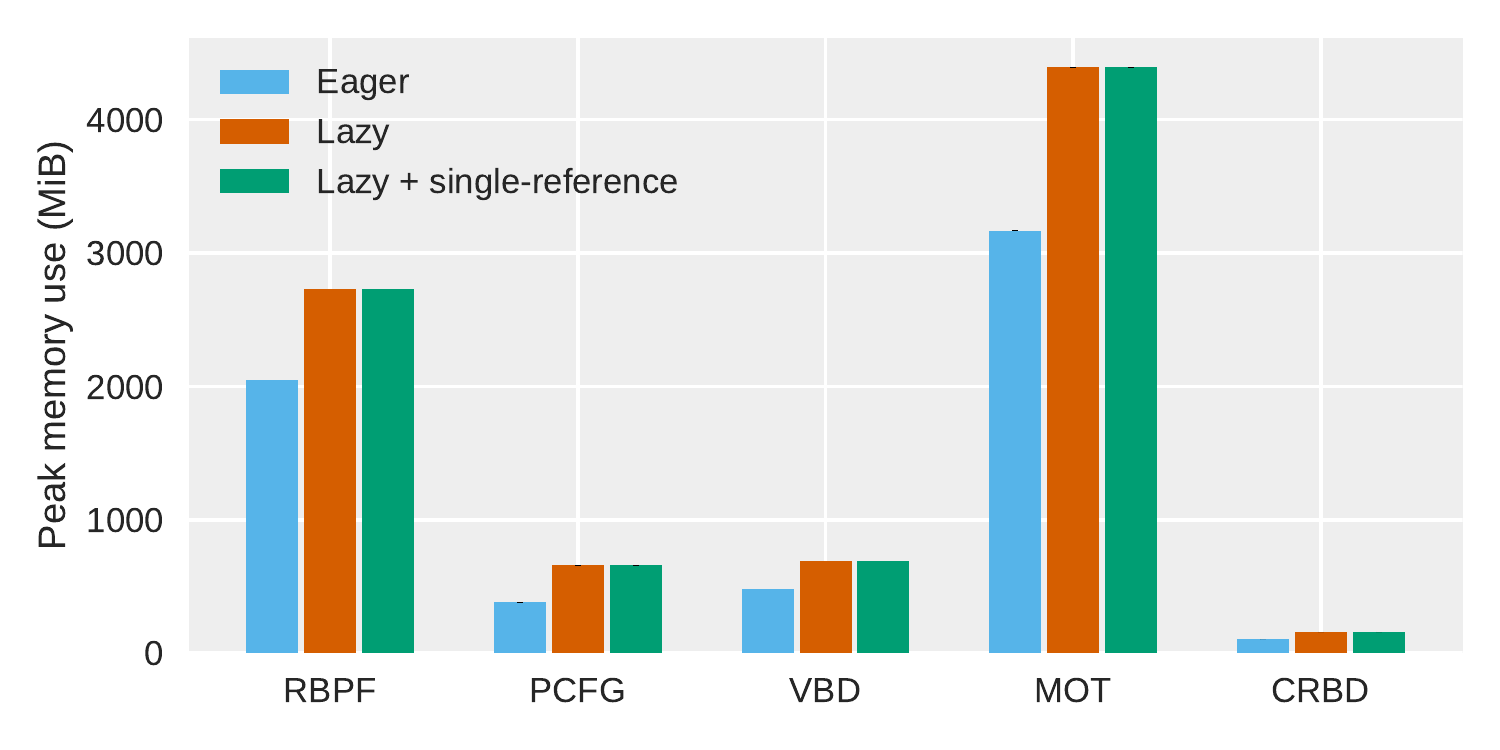}\caption{Execution time (left) and peak memory use (right) for the simulation
task. Heights indicate median, and error bars the interquartile range,
across 20 runs. This task isolates the overhead of lazy copies when
unused in a program.\label{fig:simulation}}
\end{figure*}

For the PCFG problem, the implementation of the model keeps only the
latest state, $x_{t}$, in memory, rather than the full chain, $x_{0:t-1}$.
Lazy copies are not expected to offer more than a constant factor
improvement for this, which may be outweighed by overhead. Nonetheless,
an improvement in execution time is apparent, with only a small increase
in memory use.

\begin{figure}[tp]
\includegraphics[width=0.47\textwidth]{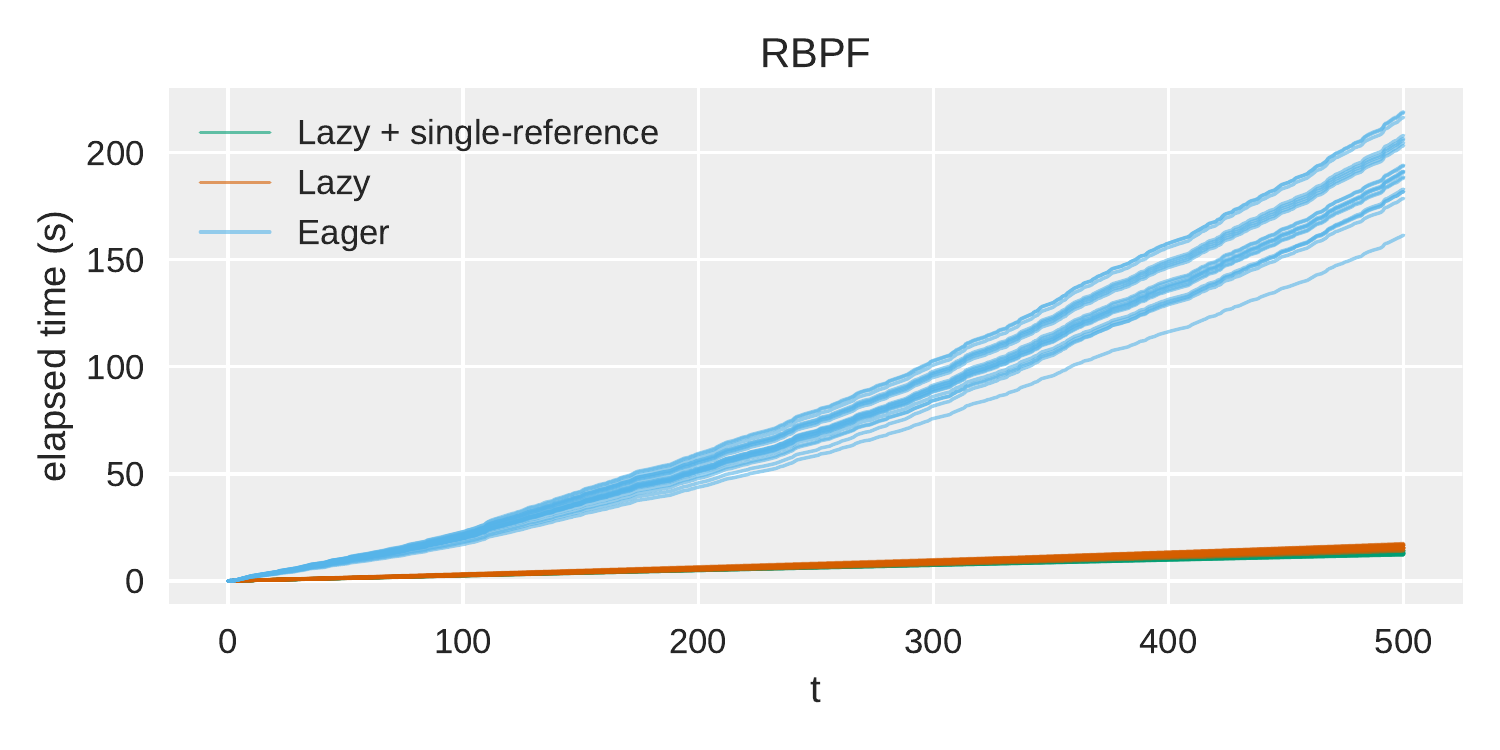}\hfill{}\includegraphics[width=0.47\textwidth]{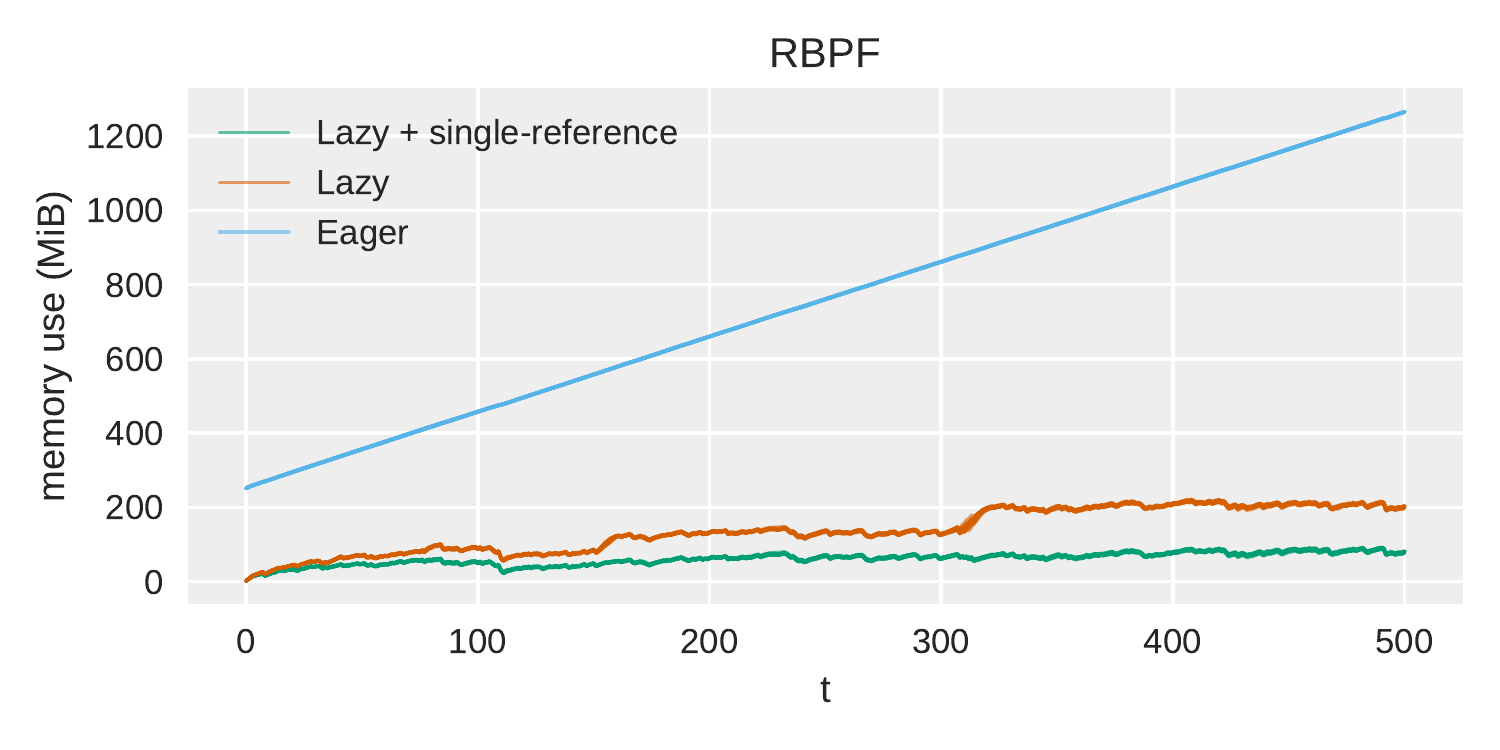}

\includegraphics[width=0.47\textwidth]{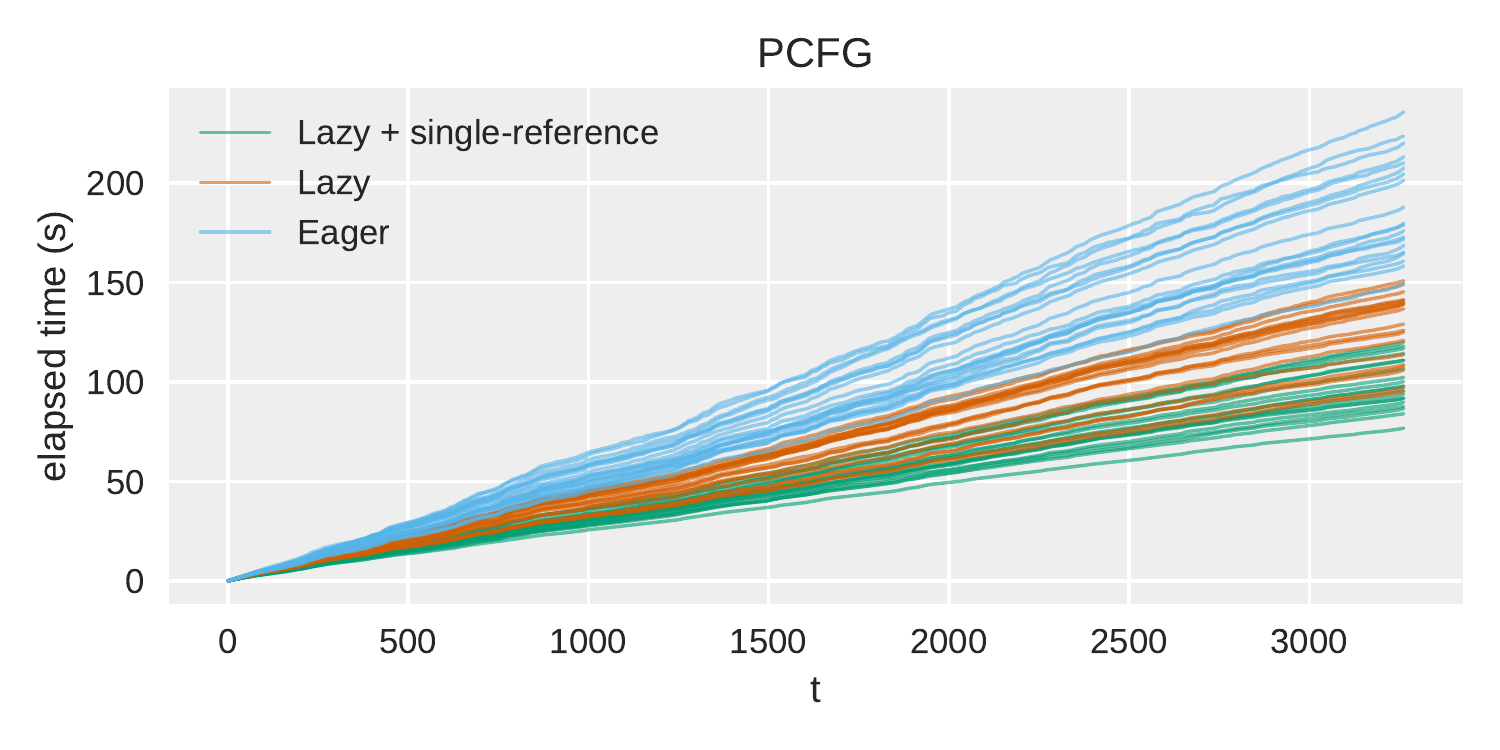}\hfill{}\includegraphics[width=0.47\textwidth]{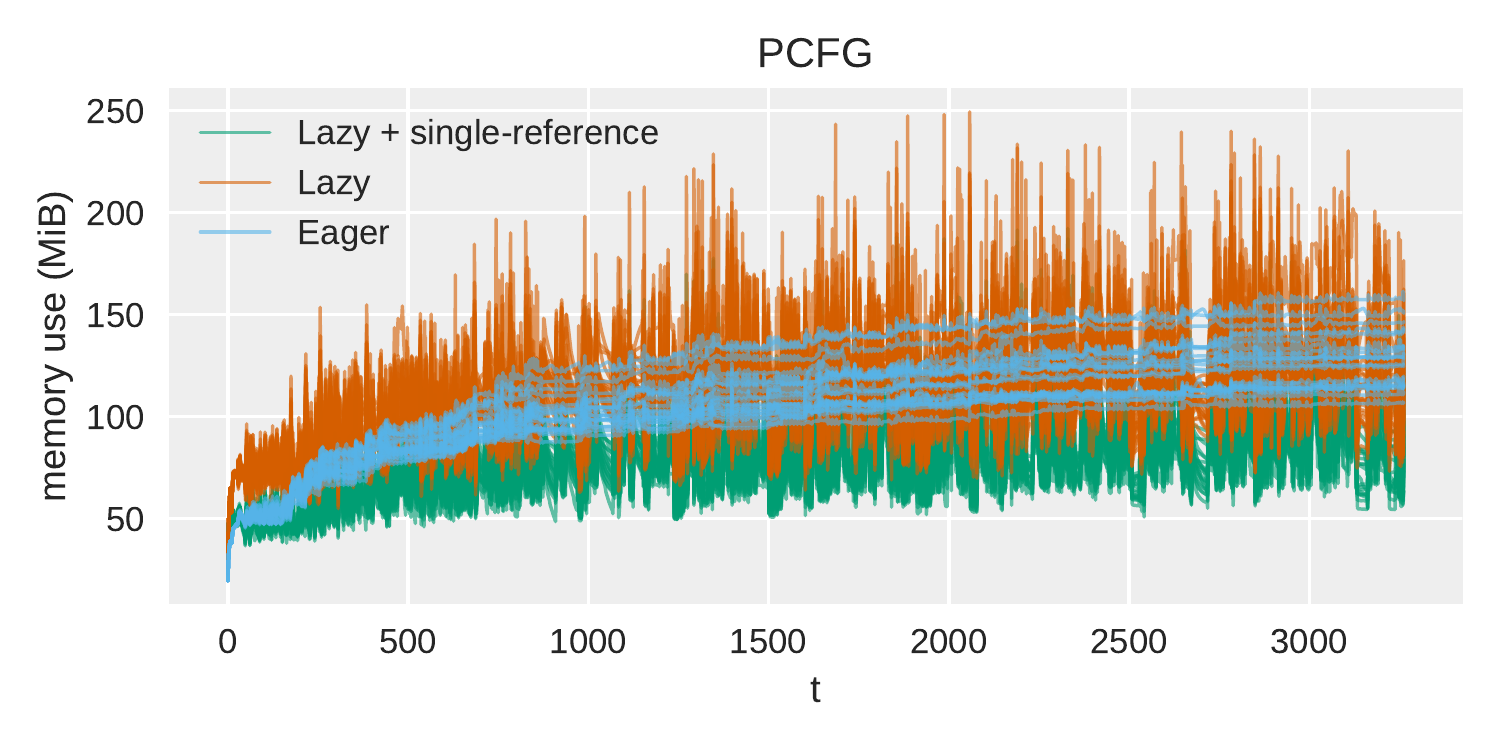}

\includegraphics[width=0.47\textwidth]{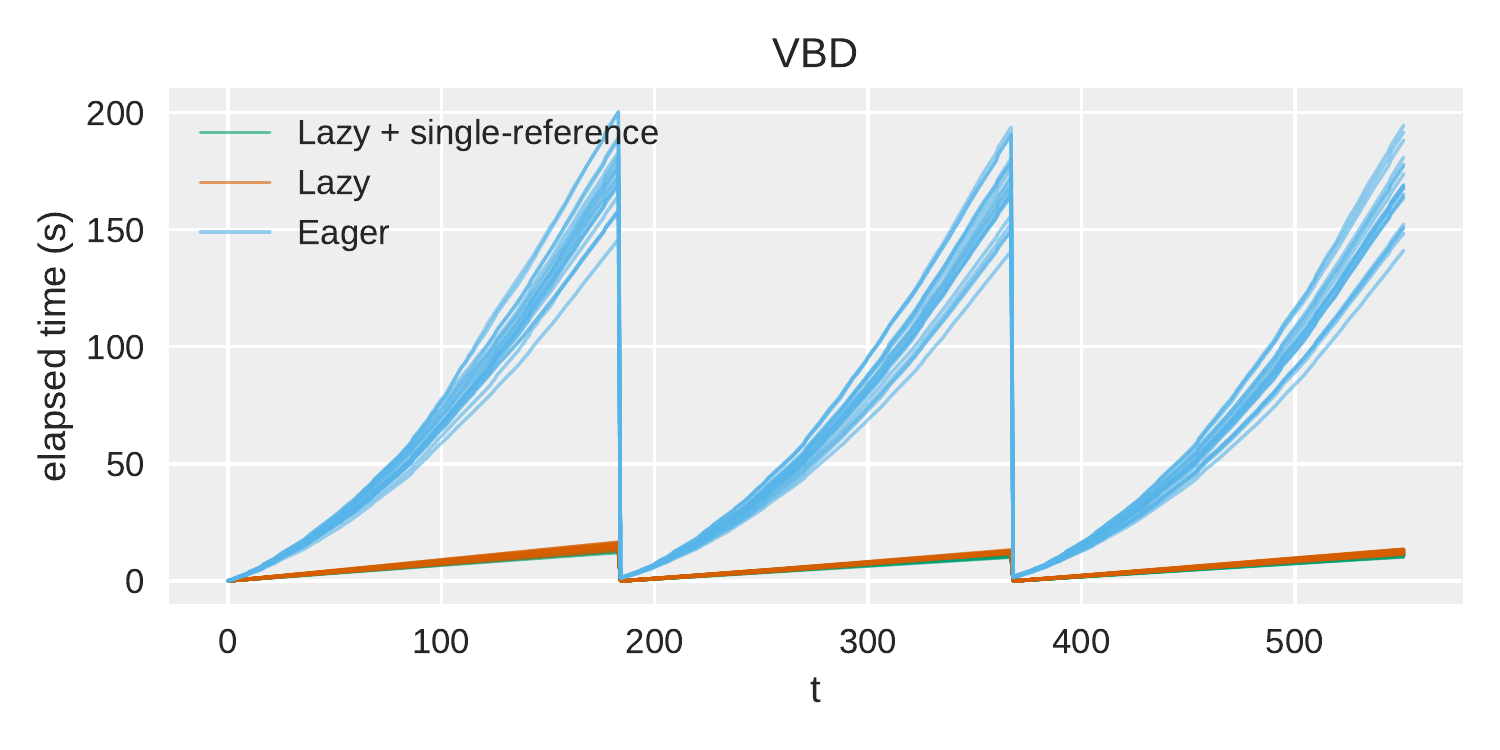}\hfill{}\includegraphics[width=0.47\textwidth]{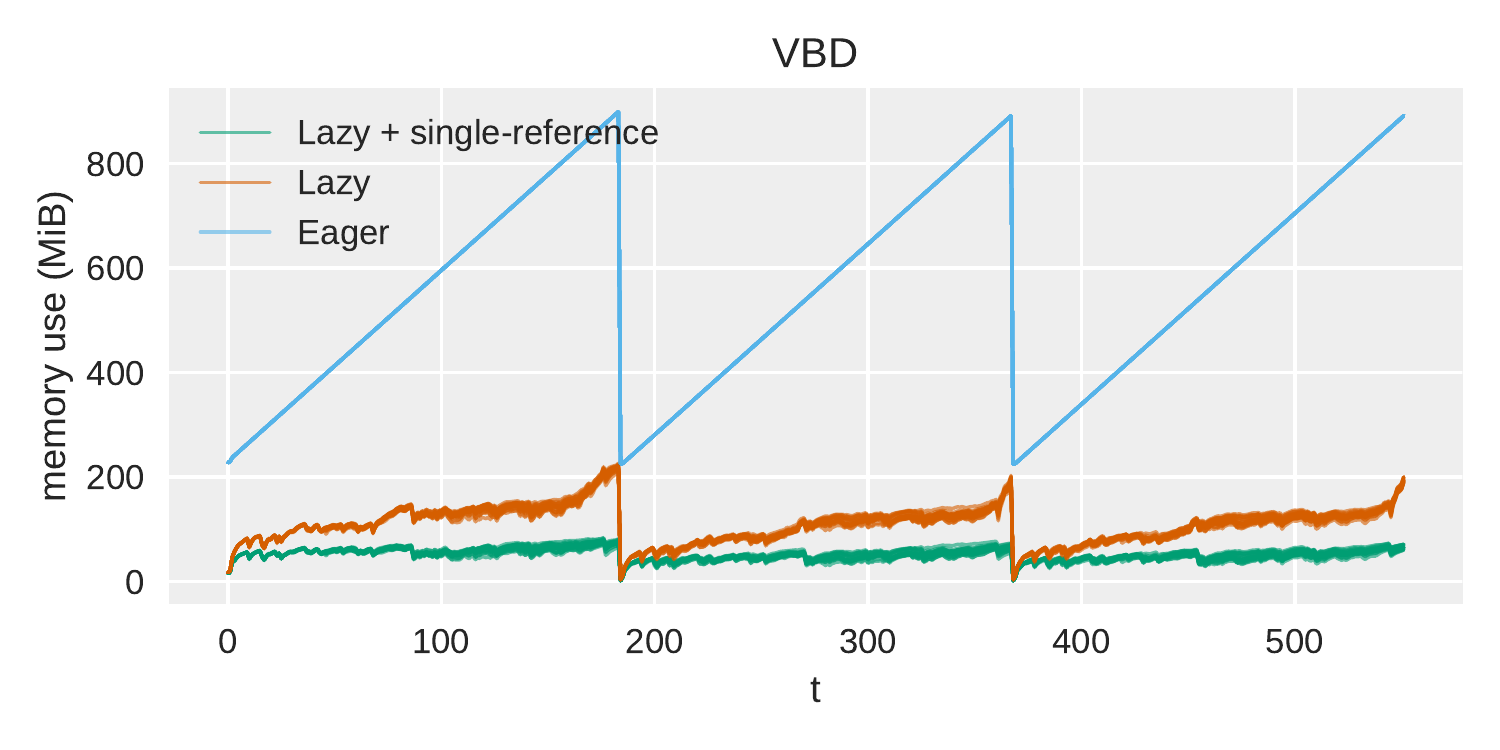}

\includegraphics[width=0.47\textwidth]{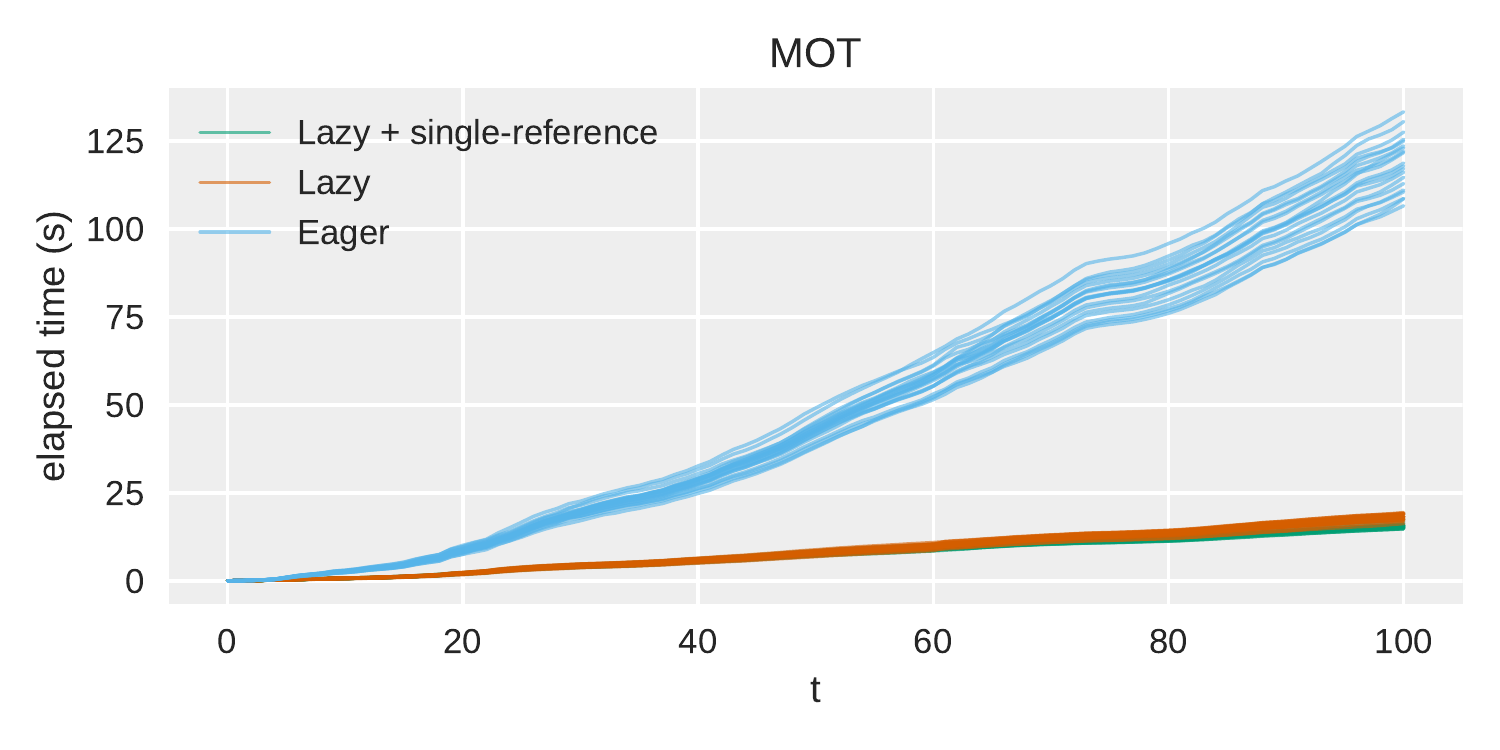}\hfill{}\includegraphics[width=0.47\textwidth]{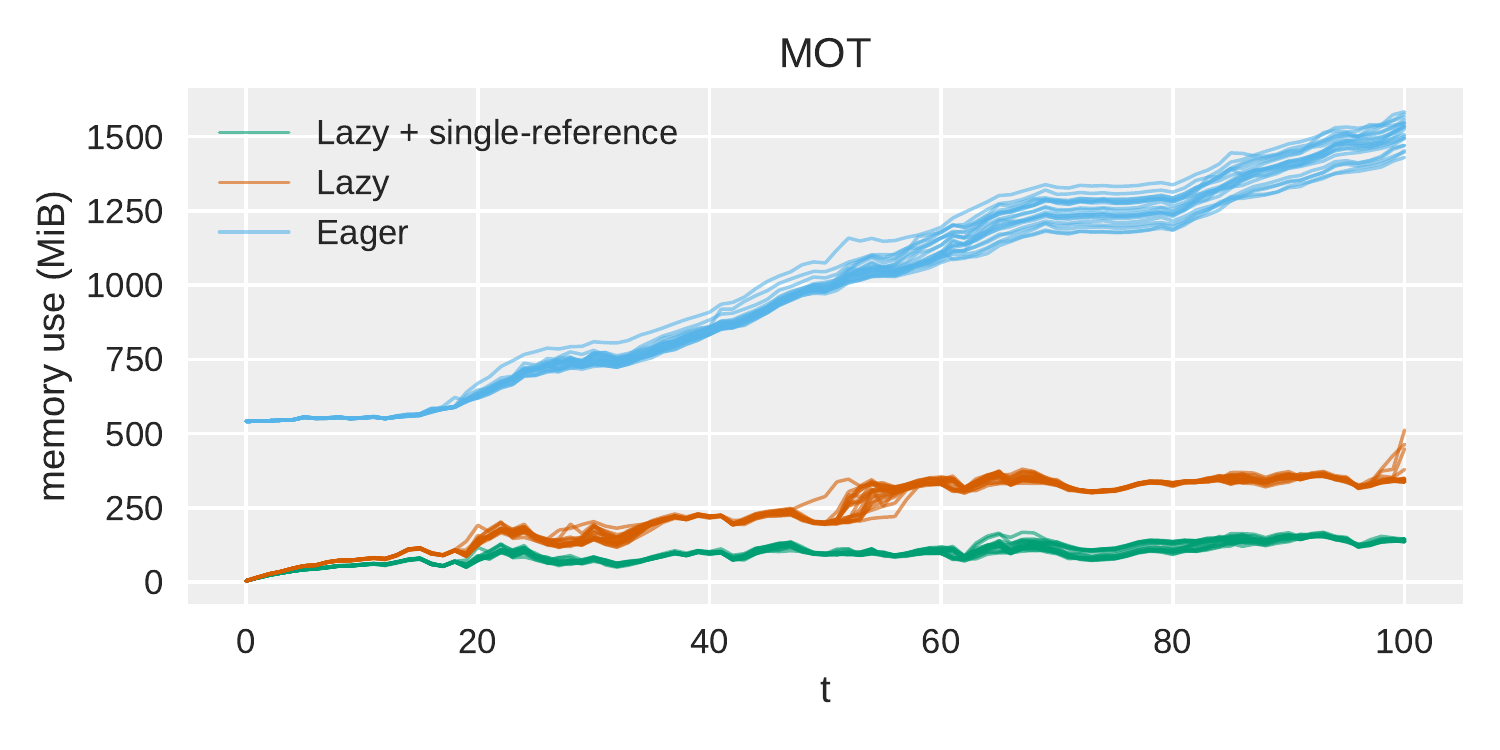}

\includegraphics[width=0.47\textwidth]{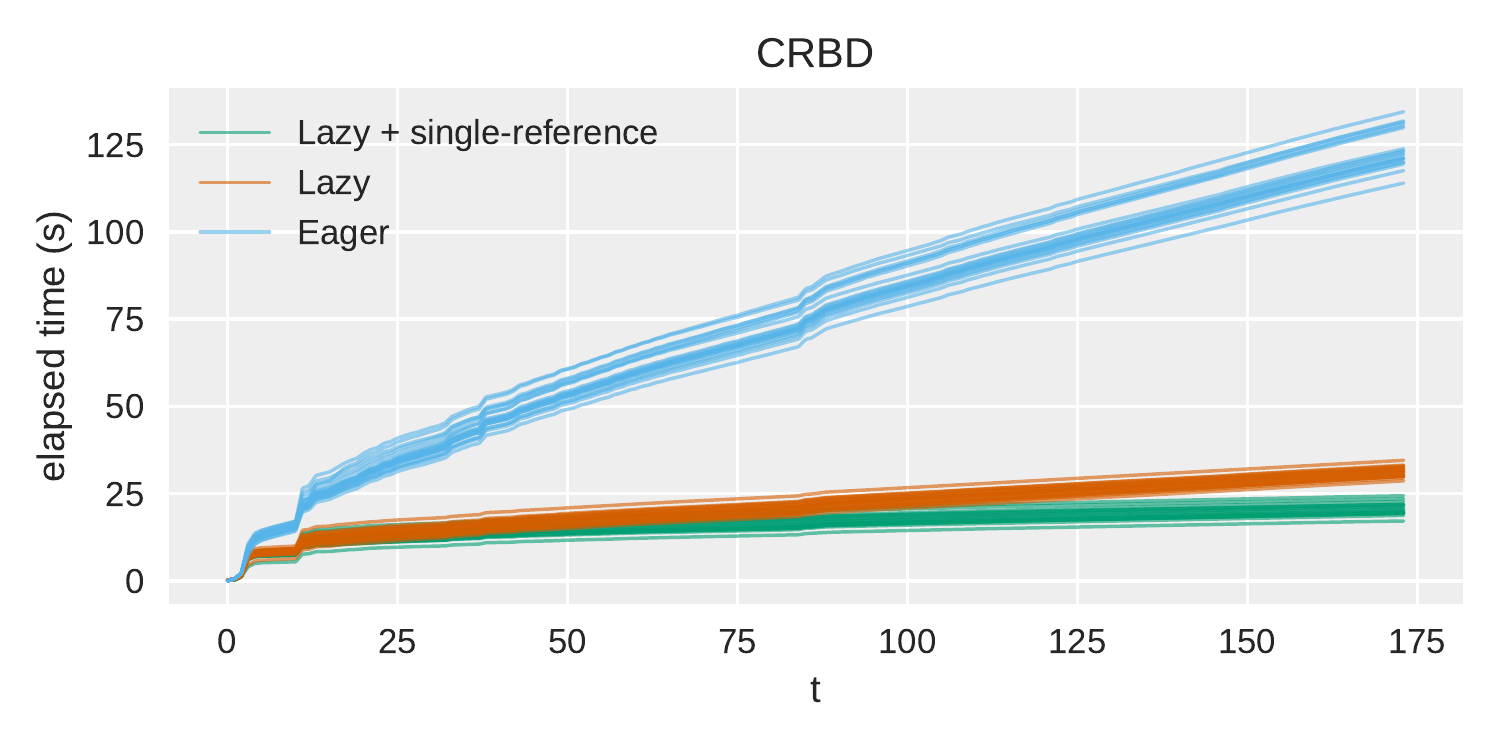}\hfill{}\includegraphics[width=0.47\textwidth]{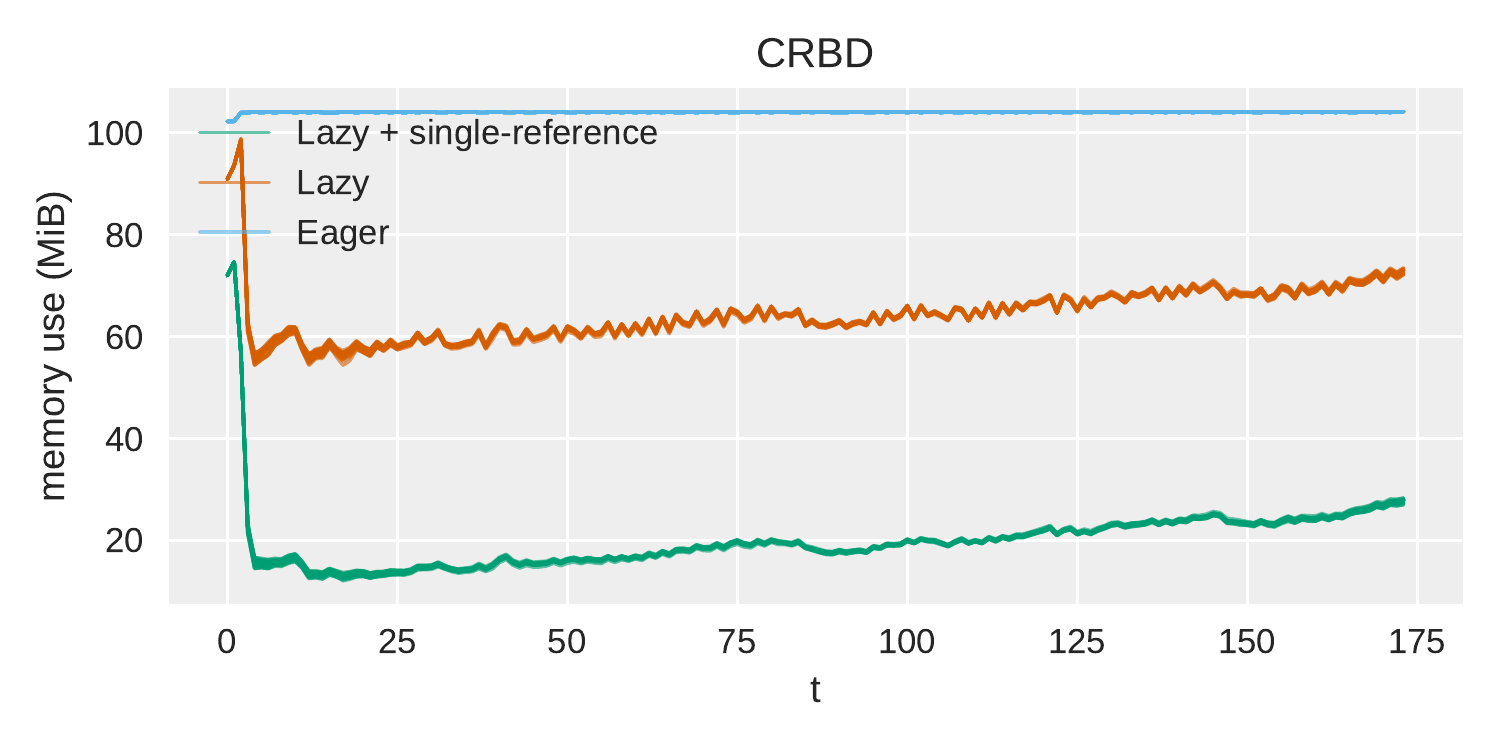}

\caption{Elapsed time and memory use across $t=1,\ldots,T$. Results are consistent
with the theoretical expectation that the eager copy configuration
should exhibit quadratic execution time and linear memory use in $t$,
while the lazy copy configurations exhibit linear execution time and
slower linear memory use in $t$, except for the PCFG example, discussed
in the text.\label{fig:over-checkpoints}}
\end{figure}

The theoretical results described in Section \ref{sec:introduction}
suggest that the ratio between the memory use of the eager and lazy
strategies will eventually increase with $t$ for all but the PCFG
and CRBD problems (for the reasons given above). Figure~\ref{fig:over-checkpoints}
shows execution time and memory use over $t$ for the inference task
on each problem. While complexity cannot be established empirically,
the results at least seem consistent with such an interpretation,
especially with the single-reference optimization enabled.

\section{Discussion and conclusion\label{sec:discussion}}

This work has considered population-based probabilistic programs and
the specialization of dynamic memory management to their particular
usage patterns. The problem and proposed solution was formalized using
labeled directed multigraphs. Whereas, in the general case, the labeling
scheme requires that each edge is assigned a list of labels, for tree-structured
copies a single label is sufficient, which significantly reduces bookkeeping.
This single label can be used to accommodate these common tree-structured
copies lazily, while handling other copies eagerly.

The key impact of the approach is enabling mutability of objects while
increasing object sharing and reducing memory use. Mutability is necessary
for imperative programming, but may also facilitate in-place write
optimizations for functional programming. It does not, of course,
alleviate the programmer from making sensible choices. Where mutability
is the sensible choice, however, the platform accommodates this; examples
include where an in-place algorithm is optimal, where contiguous storage
is required for cache efficiency, and where incremental modifications
to data structures are not well-accommodated by immutable objects.

Empirically, the approach has been shown to offer significant improvements
in terms of execution time and memory use across a number of realistic
probabilistic programs.

\section{Supplementary materials}

The platform is implemented in C++ as part of LibBirch, the support
library for the Birch probabilistic programming language, available
at \url{birch-lang.org}.

\bibliographystyle{abbrvnat}
\bibliography{lazy_deep_clone}

\end{document}